\documentclass[journal=jctcce,layout=twocolumn,manuscript=article]{achemso}
\usepackage{amsmath}
\usepackage{amssymb}
\usepackage{mathtools}
\usepackage{epsfig}
\usepackage{graphicx}
\usepackage{multicol}
\usepackage{color,soul}
\usepackage[toc,page]{appendix}
\usepackage{multirow}
\usepackage{geometry}
\usepackage{subfig}

\makeatletter
\newsavebox{\@brx}
\newcommand{\llangle}[1][]{\savebox{\@brx}{\(\m@th{#1\langle}\)}%
  \mathopen{\copy\@brx\kern-0.5\wd\@brx\usebox{\@brx}}}
\newcommand{\rrangle}[1][]{\savebox{\@brx}{\(\m@th{#1\rangle}\)}%
  \mathclose{\copy\@brx\kern-0.5\wd\@brx\usebox{\@brx}}}
\makeatother

\makeatletter
\let\l@addto@macro\relax
\makeatother
\let\oldmaketitle\maketitle
\let\maketitle\relax

\author{Sabry G. Moustafa}
\email{smoustaf@trinity.edu}
\affiliation{Department of Engineering Science, Trinity University, San Antonio, Texas 78212, USA}
\author{Andrew J. Schultz}
\affiliation{Department of Chemical and Biological Engineering, University at Buffalo, The State University of New York, Buffalo, New York 14260, USA}

\title{Generalized Path Integral Energy and Heat Capacity Estimators of Quantum Oscillators and Crystals using Harmonic Mapping}

\usepackage[fontsize=9pt]{scrextend}

\usepackage{xr}
\makeatletter

\newcommand*{\addFileDependency}[1]{
\typeout{(#1)}
\@addtofilelist{#1}
\IfFileExists{#1}{}{\typeout{No file #1.}}
}\makeatother
\newcommand*{\myexternaldocument}[1]{%
\externaldocument{#1}%
\addFileDependency{#1.tex}%
\addFileDependency{#1.aux}%
}
\myexternaldocument{si}

\begin{document}
\twocolumn[
\begin{@twocolumnfalse}
\oldmaketitle
Imaginary-time path integral (PI) is a rigorous tool to treat nuclear quantum effects in static properties. However, with its high computational demand, it is crucial to devise precise estimators. We introduce generalized PI estimators for the energy and heat capacity that utilize coordinate mapping. While it can reduce to the standard thermodynamic and centroid virial (CVir) estimators, the formulation can also take advantage of harmonic character of quantum oscillators and crystals to construct a coordinate mapping. This yields harmonically mapped averaging (HMA) estimators, with mappings that decouple (HMAc) or couple (HMAq) the centroid and internal modes. The HMAq is constructed with normal mode coordinates (HMAq-NM) with quadratic scaling of cost or harmonic oscillator staging (HMAq-SG) coordinates with linear scaling. The estimator performance is examined for a 1D anharmonic oscillator and a 3D Lennard-Jones crystal using path integral molecular dynamics (PIMD) simulation. The HMA estimators consistently provide more precise estimates compared to CVir, with the best performance obtained by HMAq-NM, followed by HMAq-SG, and then HMAc. We also examine the effect of anharmonicity (for AO), intrinsic quantumness, and Trotter number. The HMA formulation introduced assumes the availability of forces and Hessian matrix; however, an equally efficient finite difference alternative is possible when these derivatives are inaccessible. The remarkable improvement in precision offered by HMAq estimators provides a framework for efficient PI simulation of more challenging systems, such as those based on \textit{ab initio} calculations.\\
\end{@twocolumnfalse}]

\section{INTRODUCTION}
The Feynman path integral (PI) formulation of quantum mechanics provides a rigorous tool to treat nuclear quantum effects of static properties at finite temperatures.\cite{Feynman1965} These effects become increasingly significant as the thermal energy decreases relative to the spacing between quantum energy levels.\cite{ceriotti2018review,tuckerman2023book} Systems with lightweight elements (e.g., hydrogen) and stiff mode (e.g., at high pressure) are more susceptible to these effects. We focus in this study on the PI formulation of indistinguishable particles, where exchange effects are negligible and, hence, quantum Boltzmann statistics can be used.\cite{ceperley1995review} In this representation, the quantum system is mapped onto an extended classical isomorphism consisting of $n$ replicas (Trotter number, or beads) of the actual system, connected through harmonic springs. The true quantum behavior is then recovered in the limit of infinite number of beads. In the primitive PI approximation (free-article reference), the partition function of $N$ indistinguishable particles of mass $m$, at temperature $T$, is given by,\cite{tuckerman2023book}
\begin{subequations}
\label{eq:ZUeff_n_def}
\begin{align}
\label{eq:Z_n_def}
Z\left(\beta \right)  &= \left(\frac{mn}{2\pi \hbar^2 \beta}\right)^{dnN/2} \int {\rm d}^{nN} {\bf x} 
\; {\rm exp}\left(-\beta V\right),   \\
\label{eq:U_eff_def}
V \left({\bf x}, \beta\right) &= \frac{1}{n} \sum_{i=0}^{n-1} \frac{1}{2} m \omega_n^2 \left({\bf x}_i - {\bf x}_{i+1} \right)^2 +  \frac{1}{n} \sum_{i=0}^{n-1} U\left({\bf x}_i\right) \nonumber \\
&\equiv K + U
\end{align}
\end{subequations}
where $\omega_n \equiv n/\beta\hbar$, with units of angular frequency, $\hbar\equiv h/2\pi$ is the reduced Planck constant, and $\beta\equiv 1/k_{\rm B}T$, with $k_{\rm B}$ the Boltzmann constant (set to unity throughout). The vector $\bf x$ contains all bead coordinates, with ${\bf x}_{i}$ being the coordinates associated with the $i^{\rm th}$ replica, such that ${\bf x}_{n} = {\bf x}_0$. In this formulation, $V$ represents a $T$-dependent effective potential comprising intermolecular interactions $U\left({\bf x}_i\right)$, only a function of coordinates from the same replica, and a kinetic term, represented by harmonic interactions of identical spring constants, $m\omega_n^2/n$. This primitive approximation corresponds to a second-order Trotter factorization in $\beta/n$. Hence, for a given accuracy, a larger number of beads is needed at lower temperatures, such that $\beta/n$ (or, $\omega_n$) remains constant.

Given the classical isomorphism picture of quantum system, thermodynamic properties can be determined using usual statistical mechanics relations. For example, the energy $E$ and isochoric heat capacity $C_{\rm V}$ are given through derivatives of the Helmholtz free energy ($A=-k_{\rm B}T \ln Z$) with respect to $\beta$,
\begin{align}
\label{ECv_thermo}
E \equiv \frac{\partial \left(\beta A\right)}{\partial \beta}, \;\;\;\;
\frac{C_{\rm V}}{k_{\rm B} \beta^2} \equiv -\frac{\partial^2 \left(\beta A\right)}{\partial \beta^2}
\end{align}
where both derivatives are taken at constant volume $V$ and number of atoms $N$. However, compared to classical simulations, PI calculations are computationally demanding, particularly when nuclear quantum effects are pronounced (large $n$ values). This, in turn, urges continuous efforts\cite{barker1979quantum,herman1982path,parrinello-rahman1984,cao1989energy,neirotti2000heat,yamamoto2005path,martyna2007low,shiga2005Cv}   to develop efficient PI estimators that provide precise estimates using tractable computational efforts.

A naive evaluation of these free energy derivatives assumes fixed coordinates as the derivatives being taken. The approach, developed by Barker in 1979,\cite{barker1979quantum} and is known as the primitive or thermodynamic estimator, with the following expressions:   
\begin{subequations}
\label{eq:ECv_prim}
\begin{align}
\label{eq:E_prim}
E & =  \frac{dNn}{2\beta}  + \left< U-K  \right>,\\
\label{eq:Cv_prim}
\frac{ C_{\rm V}}{k_{\rm B}\beta^2}  &= 
\frac{dNn}{2\beta^2}   - \frac{2}{\beta}\left< K   \right> + {\rm var}\left(\hat E\right)
\end{align}
\end{subequations}
where $\hat E$ represents the instantaneous energy sample being averaged, such that $E=\langle \hat E\rangle$, and ${\rm var}(\hat E)\equiv\langle\hat E^2\rangle-\langle\hat E\rangle^2$ is the variance in the energy samples. The computational cost of both estimators scale linearly with the number of beads $n$, as well as the number of atoms $N$. However, due to the kinetic term $K$, the estimators are known for their substantial fluctuations. This causes the statistical uncertainty in ensemble averages to grow linearly with $n$, which is particularly problematic at low temperatures where large $n$ values are required.\cite{herman1982path} Consequently, we have opted to exclude this estimator from our analysis.

To eliminate this growth in uncertainty, Herman et al. introduced in 1982,  the virial estimator, in which they employed the virial theorem to replace the kinetic term with an equivalent, yet well-behaved, expression.\cite{herman1982path} This, in turn, yields statistical uncertainties in ensemble averages that are independent of the Trotter number. However, the estimator is limited to bound systems, such as quantum oscillators.\cite{cao1989energy,yamamoto2005path} Several attempts have been proposed to extend the applicability of the virial approach to unbound systems, such as fluids and crystals. For example, in 1984, Parrinello and Rahman\cite{parrinello-rahman1984} proposed a modified version (known also as generalized virial\cite{cao1989energy}), in which one bead is used as a reference point for the virial part. A closely related (yet, more common) estimator is the centroid virial, in which the center of mass of the ring-polymer (centroid) is used as a reference instead,\cite{fried2002-e,fried2002-cv} 
\begin{subequations}
\label{eq:ECv_cvir}
\begin{align}
\label{eq:E_cvir}
E  &=  \frac{dN}{2\beta} + \left<  U  - \frac{1}{2}  \sum_{i=0}^{n-1}  {\bf F}_i^{\rm phy} \cdot   \left( {\bf x}_i - {\bf x}_{\rm c}\right)  \right> \\
\label{eq:Cv_cvir}
\frac{C_{\rm V}}{k_{\rm B} \beta^2}   
&= \frac{dN}{2\beta^2}   + \frac{1}{4\beta} \left< 3  \sum_{i=0}^{n-1}  {\bf F}_i^{\rm phy}  \cdot  \left( {\bf x}_i - {\bf x}_{\rm c}\right) \nonumber \right. \\
&- \left.   \sum_{i=0}^{n-1} \left( {\bf x}_i - {\bf x}_{\rm c}\right) \cdot  H_{ii}^{\rm phy} \cdot \left( {\bf x}_i - {\bf x}_{\rm c}\right) \right>
+ {\rm var}\left(\hat E\right)
\end{align}
\end{subequations}
where ${\bf x}_{\rm c} \equiv \frac{1}{n} \sum_{i=0}^{n-1} {\bf x}_i$ is the centroid coordinate, and ${\bf F}_i^{\rm phy} \equiv -\frac{\partial U}{\partial {\bf x}_i}$ and $H_{ii}^{\rm phy} \equiv \frac{\partial^2 U}{\partial {\bf x}_i \partial {\bf x}_i}$ are the force vector ($dN\times 1$) and Hessian matrix ($dN\times dN$) associated with the physical potential $U$, for replica $i$. This specific heat capacity expression is known as the \textit{double} centroid virial, due to applying the same derivative associated with the energy estimator twice.\cite{fried2002-cv,shiga2005Cv,yamamoto2005path} 

These expressions suggest that, similar to the thermodynamic estimator, the computational cost of both estimators scales linearly with $n$. However, while $E$ scales linearly with the number of atoms $N$, $C_{\rm V}$ scales quadratically due to the Hessian matrix.

The centroid virial approach relies on energy derivatives (forces and Hessian), which may not be always available. For these cases, Glaesemann and Fried introduced in 2002 an alternative approach (centroid thermodynamic) that only uses the energy.\cite{fried2002-e,fried2002-cv}. In this approach, the umbrella sampling is employed to reduce fluctuations in the thermodynamic estimator. However, despite its complexity, the method did not offer improvement over the centroid virial approach, for both the energy and heat capacity. Moreover, it performs best at high temperatures, where quantum effects are less important. A closely related methodology was introduced in 2005 by Yamamoto, in the coordinate scaling corresponds to the centroid virial derivatives.\cite{yamamoto2005path} Although no new estimator was introduced, knowing the coordinate mapping associated with these estimators allowed both the energy and heat capacity to be computed from finite differences, without relying on the forces and hessian derivatives (not always accessible). This technique is particularly useful with the centroid virial estimator of the heat capacity, because it reduces the computational scaling with $N$ from quadratic to linear, due to the absence of Hessian derivatives.  

In principle, these conventional estimators are \textit{ad hoc} approaches which attempt to reduce fluctuations, either through replacing some term by more well-behaved alternative (centroid virial), or by augmenting an existing estimator by a new term (centroid thermodynamic). In other words, although they can provide reasonable level of precision in some cases, they do not offer a prescription for a systematic improvement methodology.

It is worth pointing out to an alternative PI formulation (yet, less common) which uses the HO as a reference for the PI action, rather than free-particle as the case with standard/primitive action (eq~\ref{eq:ZUeff_n_def}).\cite{friesner1984opt,martyna2007low} Although both PI versions are second-order discretization schemes, the primary advantage of this action is its smaller (or non for HO) dependence on the Trotter number. However, compared to standard PI, the formulation is more complex due to temperature-dependent of effective HO spring constants. This adds complexity to the analysis involving free energy derivative with respect to temperature, such as the energy and heat capacity we consider here. In fact, Whitfield and Martyna already developed an efficient energy estimator, using staging coordinate in these coordinates.\cite{martyna2007low} However, like standard estimators, it is an \textit{ad hoc} method designed to reduce fluctuations by introducing additional terms, which does not offer a systematic route to derive other properties (e.g., heat capacity or pressure). Moreover, generalizing this formulation (both sampling and estimator) to crystalline systems is not clear. Therefore, we rely in this study on the more simple primitive path integral formulation. As for the effect of Trotter number, we refer the reader to an efficient extrapolation scheme in standard PI, which provides accurate estimation of the continuum limit using relatively small system sizes.\cite{moustafa2024fse}

In this paper, we introduce generalized PI estimators for the total energy and heat capacity, using the mapped averaging formulation.\cite{schultz2016reformulation} The generic nature of this approach allows for recovering the standard thermodynamic and centroid virial estimators through a specific selection of the coordinates mapping field. Moreover, we have utilized this formulation to develop novel estimators for systems with harmonic character, such as quantum oscillators and crystals. The method, denoted harmonically mapped averaging (HMA), relies on the harmonic oscillator (HO) as a reference to define the mapping field. The HMA approach yields two main types of estimators: HMAc and HMAq, with mapping defined based on classical and quantum HO reference models, respectively. The HMAc and HMAq estimators are expressed in real-space coordinates. However, the mapping in the HMAq method is derived in both normal mode (HMAq-NM) and staging (HMAq-SG) coordinates. We apply this framework to a one-dimensional anharmonic oscillator and a three-dimensional Lennard-Jones crystal, at temperatures of important quantum effects. In all cases considered, the HMA approach provided remarkable improvement in precision over the centroid virial, with HMAq-NM showing the best performance, followed by HMAq-SG, and then HMAc.

The paper is organized as follows. In Section~\ref{sec:methods}, we begin by deriving generalized energy and heat capacity estimators using coordinate mapping formulation, showing that standard estimators emerge as special cases of this framework. We follow this by introducing novel estimators specific to systems with harmonic character, using harmonic mapping. Additionally, we provide details on the model and simulations employed. In Section~\ref{sec:results}, we apply these estimators to a 1D anharmonic oscillator and a 3D Lennard-Jones crystal, comparing their precision against conventional estimators at different temperatures, anharmonicity, quantumness conditions, and system sizes. We finally conclude in Section~\ref{sec:conclusions} with a summary of the methodology and its impact, discuss limitations, and explore potential future applications and extensions.

\section{FORMALISM}
\label{sec:methods}
\subsection{Generalized Path Integral Estimators}
Applying the mapped averaging formulation\cite{moustafa2015pre,schultz2016reformulation} to the PI partition function (eq~\ref{eq:Z_n_def}), using the fee energy derivatives definition of the energy and heat capacity (eq~\ref{ECv_thermo}), yields 
\begin{subequations}
\label{eq:ECv_Lag}
\begin{align}
\label{eq:ECv_Lag_E}
E &= 
\frac{dNn}{2\beta}  - \left< \sum_{i=0}^{n-1}\nabla \cdot \dot {\bf x}_i \right>  + \left<  \frac{D \left(\beta V\right)}{D \beta} \right> ,\\
\label{eq:ECv_Lag_Cv}
\frac{C_{\rm V}}{k_{\rm B} \beta^2}  &= 
\frac{dNn}{2\beta^2}     + \left< \sum_{i=0}^{n-1} \frac{\rm D}{{\rm D}\beta}\nabla\cdot \dot {\bf x}_i \right>  
- \left< \frac{D^2\left( \beta V\right)}{D \beta^2}
\right> + {\rm Var}\left(\hat E\right)
\end{align}
\end{subequations}
where $\dot {\bf x}_i \equiv {\rm D} {\bf x}_i\left(\beta\right)/{\rm D}\beta$ is the coordinates mapping velocity with respect to $\beta$ (analogous to time) and $D/D\beta$ denotes a Lagrangian derivative with respect to $\beta$, i.e. a total derivative while scaling coordinates according to the mapping field.

With this freedom in choosing the mapping, we define it based on a reference model of exact solution, such as the ideal gas or harmonic oscillator models. In this case, all samples that go into eq~\ref{eq:ECv_Lag} are identical (zero fluctuations), such that we can remove the angle brackets. The mapping field can then be derived from the energy estimator of the energy,
\begin{align}
\label{eq:mapping_0}
E^{\rm ref}  = \frac{dNn}{2\beta}  -  \sum_{i=0}^{n-1}\nabla \cdot \dot {\bf x}_i   +   \frac{D \left(\beta V^{\rm ref}\right)}{D \beta} = {\rm constant}
\end{align}
Given that $E^{\rm ref}$ is an exactly known constant, this differential equation can be solved (in principle) for $\dot{\bf x}_i$. We show below that different choices for the reference model yield the standard and new PI estimators. It is worthwhile to emphasize that, mathematically, the free energy derivatives ($E$ and $C_{\rm V}$) depend only on the thermodynamic state (e.g., $T$) and not on the choice of the mapping field. However, the \textit{precision} of measured ensemble averages is affected, as coordinates scaling directly influences sample fluctuations. 

An important special case of mapping is the linear field (i.e., $\dot{\bf x}_i \propto {\bf x}_i$), which is the type of mapping associated with all estimators considered here, as we show below. Hence, the divergence term, $\nabla\cdot\dot {\bf x}_i$, becomes configuration-independent (still a function of $\beta$). However, since $V^{\rm ref}$ is a configuration-dependent function, its total derivative must vanish, in order for eq~\ref{eq:mapping_0} to be configuration-independent. Accordingly, eq~\ref{eq:mapping_0} reduces to $E^{\rm ref}=\frac{dNn}{2\beta}  -  \sum_{i=0}^{n-1}\nabla \cdot \dot {\bf x}_i$. Since the divergence term is configuration-independent, its ensemble average does not depend on the sampling model, such that $\langle \nabla\cdot\dot {\bf x}_i \rangle$ can be replaced by $\nabla\cdot\dot {\bf x}_i$ in eq~\ref{eq:ECv_Lag}, to yield
\begin{subequations}
\label{eq:ECv_Lag_ref}
\begin{align}
\label{eq:ECv_Lag_E_ref}
E  &= E^{\rm ref}  + \left<  \frac{D \left(\beta V\right)}{D \beta} \right> ,\\
\label{eq:ECv_Lag_Cv_ref}
\frac{C_{\rm V}}{k_{\rm B} \beta^2}  &= 
\frac{C^{\rm ref}_{\rm V}}{k_{\rm B} \beta^2}
- \left< \frac{D^2\left( \beta V\right)}{D \beta^2}
\right> + {\rm Var}\left(\hat E\right)
\end{align}
\end{subequations}
where $E^{\rm ref}$ is given above and $C_{\rm V}^{\rm ref} \equiv \frac{dNnk_{\rm B}}{2}     + k_{\rm B} \beta^2\sum_{i=0}^{n-1} \frac{\rm D}{{\rm D}\beta}\nabla\cdot \dot {\bf x}_i$. Equivalently, $E^{\rm ref}$ and $C_{\rm V}^{\rm ref}$ can be obtained directly from free energy derivatives of reference free energy.

Given the mapping velocity from eq~\ref{eq:mapping_0}, the Lagrangian derivatives of $\beta V\left({\bf x}\left(\beta\right), \beta\right)$ can be computed either numerically using finite differences, or analytically by transforming to the Eulerian representation.\cite{schultz2016reformulation} The former is most suitable when the first (forces) and second (Hessian) derivatives of the physical potential energy ($U$) are not available, such as with some \textit{ab initio} calculations. For our case, these derivatives are already available and, hence, we adopt the Eulerian approach in the next section. Nevertheless, for the sake of completeness, we provide in Section S1 of the Supporting Information detailed analysis using the finite difference approach. It is worthwhile to highlight that, although approximate, the finite difference approach yields statistically indistinguishable results from the exact Eulerian derivatives (results not shown here). This is attributed to the nearly linear variation of $\beta V\left({\bf x}\left(\beta\right)\right)$ function with respect to $\beta$.

\subsubsection{Eulerian Representation}
Using the Eulerian representation\cite{schultz2016reformulation} of the total derivatives in eq~\ref{eq:ECv_Lag} yields
\begin{subequations}
\label{eq:ECv_ma_def}
\begin{align}
\label{eq:E_ma_def}
 \frac{D \left(\beta V\right)}{D \beta}  &=   \frac{\partial \left(\beta V\right)}{\partial \beta} 
-  \beta \sum_{i=0}^{n-1}  {\bf F}_i  \cdot  \dot {\bf x}_i 
\\
\label{eq:Cv_ma_def}
\frac{D^2\left( \beta V\right)}{D \beta^2} &=   \frac{\partial^2 \left(\beta V\right)}{\partial \beta^2} 
-  \beta \sum_{i=0}^{n-1}    {\bf F}_i \cdot  {\ddot {\bf x}_i} \nonumber \\
&- 2  \sum_{i=0}^{n-1}  \frac{\partial  \left(\beta {\bf F}_i\right)}{\partial \beta} \cdot \dot {\bf x}_i + \beta \sum_{i=0}^{n-1}\sum_{j=0}^{n-1} \  \dot {\bf x}_i \cdot H_{ij} \cdot \dot {\bf x}_j
\end{align}
\end{subequations}
where ${\bf F}_i=-\frac{\partial V}{\partial {\bf x}_i}$ is the effective forces vector ($dN\times 1$) acting on beads of replica $i$,  $H_{ij}=\frac{\partial^2 V}{\partial {\bf x}_i \partial {\bf x}_j}$ is the effective Hessian matrix ($dN\times dN$) associated with replicas $i$ and $j$, and $\ddot {\bf x}_i= d^2{\bf x}_i\left(\beta\right)/d\beta^2$ is the mapping acceleration. Using the definition of effective potential $V$ in eq~\ref{eq:U_eff_def}, its derivatives are given by:
\begin{subequations}
\label{eq:dVs}
\begin{align}
\frac{\partial\left(\beta V\right)}{\partial \beta} 
&= U-K \\
\frac{\partial^2 \left(\beta V\right)}{\partial \beta^2} 
&= \frac{2}{\beta} K \\
 {\bf F}_i 
&= -\frac{\partial V}{\partial {\bf x}_i} = {\bf F}^{\rm kin}_i + {\bf F}^{\rm phy}_i   \\
H_{ij} 
&= \frac{\partial^2 V}{\partial {\bf x}_i \partial {\bf x}_j} = H_{ij}^{\rm kin} + H_{ij}^{\rm phy} \nonumber \\
&= \frac{m \omega_n^2}{n} \left(2 \delta_{i,j} I - \delta_{i-1,j} I -  \delta_{i+1,j} I \right)   + \delta_{ij}  H_{ii}^{\rm phy} \\
\frac{\partial \left(\beta {\bf F}_i\right)}{\partial \beta} 
&=  {\bf F}^{\rm phy}_i - {\bf F}^{\rm kin}_i
\end{align}
\end{subequations} 
where $I$ is the identity ($dN\times dN$) matrix, $\delta_{i,j}$ is the Kronecker delta function, and the ``kin'' and ``phy'' superscripts denote the kinetic and physical potential contributions, respectively, which are given by:   
\begin{subequations}
\label{eq:FH}
\begin{align}
{\bf F}^{\rm kin}_i &\equiv -\frac{\partial K}{\partial {\bf x}_i} = - \frac{m \omega_n^2}{n} \left(2{\bf x}_i - {\bf x}_{i-1} - {\bf x}_{i+1} \right) \\
{\bf F}_i^{\rm phy} &\equiv -\frac{\partial U}{\partial {\bf x}_i} = -\frac{1}{n}\frac{\partial U\left({\bf x}_i\right)}{\partial {\bf x}_i}\\
H_{ii}^{\rm phy} &\equiv \frac{\partial^2 U}{\partial {\bf x}_i \partial {\bf x}_i} = \frac{1}{n} \frac{\partial^2 U\left({\bf x}_i\right)}{\partial {\bf x}_i \partial {\bf x}_i}
\end{align}
\end{subequations}

Eventually, substituting eq~\ref{eq:FH} into eq~\ref{eq:dVs}, and using that in eq~\ref{eq:ECv_ma_def} yields
\begin{subequations}
\label{eq:ECv_ma}
\begin{align}
\label{eq:E_ma}
E &= E^{\rm ref} + \left< U-K
- \beta \sum_{i=0}^{n-1} {\bf F}_i \cdot  \dot {\bf x}_i \right>   \\
\label{eq:Cv_ma}
\frac{ C_{\rm V}}{k_{\rm B} \beta^2}  
&= \frac{ C^{\rm ref}_{\rm V}}{k_{\rm B} \beta^2}  + \left<-\frac{2}{\beta} K + \beta \sum_{i=0}^{n-1} {\bf F}_i  \cdot {\ddot {\bf x}_i} 
 +  2  \sum_{i=0}^{n-1}  \left( {\bf F}^{\rm phy}_i - {\bf F}^{\rm kin}_i \right)  \cdot \dot {\bf x}_i \nonumber \right. \\
&- \left.   \beta  \sum_{i=0}^{n-1}   m \omega_n^2 \left(\dot {\bf x}_i  - \dot {\bf x}_{i+1} \right)^2
- \beta \sum_{i=0}^{n-1}    \dot {\bf x}_i \cdot  H_{ii}^{\rm phy} \cdot \dot {\bf x}_i \right>  + {\rm var}\left(\hat E\right)
\end{align}
\end{subequations}
These estimators, along with the Lagrangian alternative (eq~\ref{eq:ECv_Lag_ref}), are the main equations of this work, which represent generalized PI estimators for a given mapping velocity field.

\begin{figure}
\includegraphics[width=0.356\textwidth]{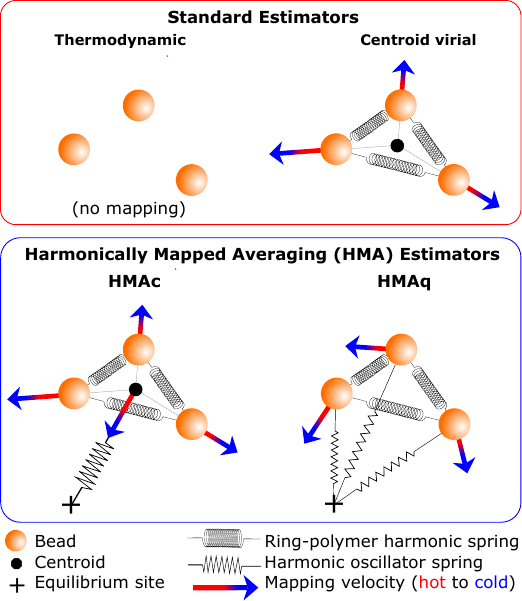}
\caption{\label{fig:mapping} Mapping velocity fields of the standard and HMA-based estimators.}
\centering
\end{figure}

\subsection{Standard Estimators}
\label{sec:standard_estimators}
We derive the mapping fields associated with two standard estimators; namely, the thermodynamic (or, Barker\cite{barker1979quantum}) and centroid virial\cite{herman1982path} estimators. Apart from the superior performance (in terms of precision) of the latter in comparison to the former, both estimators are applicable to any phase of the system.

\subsubsection{Thermodynamic Estimator}
The mapping reference model associated with the thermodynamic estimator\cite{barker1979quantum} is a \textit{classical} ideal gas (IGc) of $Nn$ noninteracting classical particles. Substituting $V=0$ into eq~\ref{eq:Z_n_def} yields
\begin{subequations}
\begin{align}
\beta A^{\rm IGc} &= \frac{dNn}{2} \ln \left(\frac{2\pi \hbar^2 \beta}{mn}\right)    
- Nn \ln \Omega, \\
E^{\rm IGc} &= \frac{dNn}{2\beta} 
\end{align}
\end{subequations}
where $\Omega$ is the volume. Therefore, the associated mapping equation (eq~\ref{eq:mapping_0}) becomes
\begin{align}
E^{\rm IGc} = \frac{dNn}{2\beta} = 
\frac{dNn}{2\beta}  -  \sum_{i=0}^{n-1}\nabla\cdot \dot {\bf x}_i  
\end{align}
The solution of this differential equation is simply $\dot {\bf x}_i = 0$ and, consequently, $\ddot {\bf x}_i = 0$. This corresponds to no coordinate mapping as (see Fig.~\ref{fig:mapping}). Using this fact in the generalized PI estimator expressions (eq~\ref{eq:ECv_ma}) yields the standard thermodynamic estimators of the energy and heat capacity (eq~\ref{eq:ECv_prim}).

\subsubsection{Centroid Virial Estimator}
\label{sec:cvir_estimator}
The mapping reference model associated with the centroid virial estimator is a \textit{quantum} ideal gas (IGq) system of $N$ noninteracting particles. Substituting $U=0$ into eq~\ref{eq:Z_n_def} yields
  \begin{subequations}  
\begin{align}
\beta A^{\rm IGq}  &= \frac{dN}{2} \ln \left(\frac{2\pi \hbar^2 \beta}{m}  \right) - N \ln \Omega,\\
E^{\rm IGq} &\equiv \frac{\partial \left(\beta A^{\rm IGq}\right)}{\partial \beta}= \frac{dN}{2\beta} 
\end{align}
\end{subequations}
Using this energy expression, along with $V^{\rm IGq}=K$, in eq~\ref{eq:mapping_0}, yields the following mapping equation,
\begin{align}
\frac{dN}{2\beta} &=  \frac{dNn}{2\beta}  -  \sum_{i=0}^{n-1}\nabla\cdot \dot {\bf x}_i 
- \frac{1}{n} \sum_{i=0}^{n-1} \frac{1}{2} m \omega_n^2 \left({\bf x}_i - {\bf x}_{i+1} \right)^2  \nonumber\\
&+\frac{\beta}{n} \sum_{i=0}^{n-1} \left(  m \omega_n^2 \left(2{\bf x}_i - {\bf x}_{i-1} - {\bf x}_{i+1} \right) \right)  \cdot  {\dot {\bf x}_i}
\end{align}
The solution for this differential equation, along with its first derivative (acceleration), are given by
\begin{subequations}
\label{eq:mapping_cv}
\begin{align}
\label{eq:xdot_cv}
\dot{\bf x}_i &= \frac{1}{2\beta}\left( {\bf x}_i-{\bf x}_{\rm c}\right),\\
\label{eq:xddot_cv}
\ddot{\bf x}_i &=   -\frac{1}{4\beta^2} \left( {\bf x}_i-{\bf x}_{\rm c}\right)
\end{align}
\end{subequations}
The acceleration expression is obtained by taking the total derivative of eq~\ref{eq:xdot_cv} ($\ddot{\bf x}_i  = \frac{{\rm D}\dot{\bf x}_i}{{\rm D}\beta}$) and recognizing the fact that $\dot {\bf x}_{\rm c}=0$ for this specific mapping, because summing eq~\ref{eq:xdot_cv} over $i$ is zero. This mapping field is represented graphically in Fig.~\ref{fig:mapping}. Substituting this mapping field, along with ${\bf F}^{\rm phy}_i =0$ and $H_{ii}^{\rm phys}=0$, into eq~\ref{eq:ECv_ma} results in the standard  expressions of the centroid virial estimator (eq~\ref{eq:ECv_cvir}). 

As mentioned earlier, the centroid virial expression for the heat capacity is also known as the double centroid virial.\cite{fried2002-cv,shiga2005Cv,yamamoto2005path} This could be now understood in the context of mapped averaging as we apply the same centroid virial mapping field twice. However, a mixed estimator could also be derived.\cite{fried2002-cv,shiga2005Cv} The mapping in this case will be a mix between two different mapping methods; for example using the centroid virial for the first derivative, followed by a second derivative using the thermodynamic mapping. Although these mixed estimators are Hessian-free, they are less precise that the double centroid mapping approach.\cite{shiga2005Cv}

\subsection{Harmonically Mapped Averaging (HMA) Estimators}
\label{sec:hma0}
Here, we introduce two novel estimators (HMAc and HMAq) specific to systems with harmonic character, such as quantum oscillators (e.g., molecular bonds) and crystals. Both estimators use the harmonic oscillator (or Einstein crystal for solids) as a mapping reference, which has an exact solution. With this in mind, these HMA estimators are \textit{not} applicable to fluids. 

Before proceeding, we point out to two points specific to applying HMA estimators to translationally invariant (unbound) models, such as LJ. First, a center of mass contribution associated with the HO reference model needs to be removed from HMA expressions. As detailed in Section S3 of the Supporting Information, these terms are $\frac{d}{2\beta}$ for $E$ and $\frac{d}{2\beta^2}$ for $C_{\rm V}$, which are merely finite-size effects. Hence, it is only useful to include them when comparing to standard estimators, as the case here. Second, due to the stochastic nature of the Langevin thermostat, the center of mass of the \textit{actual} system moves freely over time. This has no effect on a standard estimator like the centroid virial, because it depends only on relative positions (${\bf x}_i-{\bf x}_{\rm c}$ in this case). In contrast, as we see below, HMA estimators depends on the displacement from lattice sites. Hence for efficient performance, we subtract the center of mass contribution from all beads displacements, such that total displacement stays zero.

\subsubsection{HMAc Estimator}
\label{sec:hmac_estimator}
The mapping reference in this case is a \textit{classical} harmonic oscillator (HOc), or, equivalently, a classical Einstein crystal for solids (see Fig.~\ref{fig:mapping}). In this case, a harmonic spring is tethered (hypothetically) to the centroid of the ring-polymer (see Fig.~\ref{fig:mapping}). The associated potential energy is given by $U^{\rm  HOc} =  \frac{1}{2} m \omega^2 {\bf x}_{\rm c}^2$, which can be plugged into eq~\ref{eq:ZUeff_n_def} to yield
\begin{subequations}
\label{eq:ae_hmac}
\begin{align}
\label{eq:a_hmac}
\beta A^{\rm HOc} &= d N \ln\left( \beta \hbar \omega \right) \\
\label{eq:e_hmac}
E^{\rm HOc} &= \frac{dN}{\beta}
\end{align}
\end{subequations}
where $\omega$ is the angular frequency of the HO and $m \omega^2$ is its force constant. Following the same procedure as in Section~\ref{sec:standard_estimators}, along with the fact that ${\bf F}_i^{\rm phy}=-\frac{m\omega^2}{n} {\bf x}_{\rm c}$ for this model, the mapping velocity can be obtained from eq~\ref{eq:mapping_0} as
\begin{align}
\frac{dN}{\beta} &=  \frac{dNn}{2\beta}  -  \sum_{i=0}^{n-1}\nabla\cdot \dot {\bf x}_i 
- \frac{1}{n} \sum_{i=0}^{n-1} \frac{1}{2} m \omega_n^2 \left({\bf x}_i - {\bf x}_{i+1} \right)^2  \nonumber\\
&+\frac{\beta}{n} \sum_{i=0}^{n-1} \left(  m \omega_n^2 \left(2{\bf x}_i - {\bf x}_{i-1} - {\bf x}_{i+1} \right) \right)  \cdot  {\dot {\bf x}_i}
\end{align}
Here, ${\bf x}_i$ represent the distance vector from the equilibrium position, which is the point at which the springs are tethered for the case of oscillators, or the lattice site for the case of crystals. It follows that the solution for this differential equation, and associated acceleration, are given by
\begin{subequations}
\label{eq:mapping_hmac}
\begin{align}
\label{eq:mapping_hmac_E}
\dot {\bf x}_i &= \frac{1}{2\beta}\left(  {\bf x}_i - 2 {\bf x}_{\rm c} \right),\\
\label{eq:mapping_hmac_Cv}
 \ddot{\bf x}_i &=  - \frac{1}{4\beta^2}  \left( {\bf x}_i -4{\bf x}_{\rm c}\right)
\end{align}
\end{subequations}
where ${\bf x}_{\rm c}$ is the centroid displacement from the equilibrium position. The acceleration expression was derived using the fact that $\dot{\bf x}_{\rm c}=-\frac{1}{2\beta}{\bf x}_{\rm c}$ (sum over eq~\ref{eq:mapping_hmac_E}). This mapping field is shown in Fig~\ref{fig:mapping}, where the mapping is composed of two (independent) contribution, one from the ring-polymer (same as centroid virial) and another from the HO spring tethered to the centroid. 

Plugging this mapping field into eq~\ref{eq:ECv_ma} and simplifying, yields
\begin{subequations}
\label{eq:En_Cvn_HMAc}
\begin{align}
E &= E^{\rm HOc}  
+ \left< U 
- \frac{1}{2}\sum_{i=0}^{n-1}  {\bf F}^{\rm phy}_i  \cdot \left(  {\bf x}_i - 2 {\bf x}_{\rm c} \right)  \right>   \\
\frac{C_{\rm V}}{k_{\rm B}\beta^2}    &= \frac{C_{\rm V}^{\rm HOc}}{k_{\rm B}\beta^2}
+ \frac{1}{4\beta}\left< 3  \sum_{i=0}^{n-1}  {\bf F}^{\rm phy}_i  \cdot  \left( {\bf x}_i - 2{\bf x}_{\rm c}\right)  
+ 2 \sum_{i=0}^{n-1}  {\bf F}^{\rm phy}_i  \cdot  {\bf x}_{\rm c} \nonumber \right. \\
&- \left.   \sum_{i=0}^{n-1} \left( {\bf x}_i -2 {\bf x}_{\rm c}\right) \cdot  H^{\rm phy}_{ii} \cdot \left( {\bf x}_i - 2{\bf x}_{\rm c}\right)
\right> +  {\rm var}\left(\hat E\right)  
\end{align}
\end{subequations}
where $E^{\rm HOc}$ is given by eq~\ref{eq:e_hmac}, from which $C_{\rm V}^{\rm HOc}=dN k_{\rm B}$. As can be recognized, these expressions have similar structure to those of the centroid virial, which results in similar computational costs. However, as we will show below, the HMAc estimator provides more precise estimates. 

It can be shown that these expressions reduce to the reference model counterparts (as they should) when modelling HOc system. In this case, the associated Hessian of this model is given by $H^{\rm phy}_{ii}=\frac{m\omega^2}{n^2} I$, where $I$ is the identity matrix ($d\times d$ for HOc or $dN\times dN$ for Einstein crystal). Moreover, for this classical model we have $\left<x_i^2\right> = \left<x_{\rm c}^2\right>$, which follows from the fact that the HOc model is only function of the centroid degree of freedom. Plugging these identities into eq~\ref{eq:En_Cvn_HMAc}, and using the fact that both the forces and Hessian are independent of the index $i$, we recover the reference model properties.

\subsubsection{HMAq Estimators}
\label{sec:hmaq_estimator}
Here, we adopt the \textit{quantum} harmonic oscillator (HOq), or Einstein crystal for solids, as a reference for the mapping (see Fig.~\ref{fig:mapping}). For simplicity of notation, we will consider a single one-dimentional oscillator to derive the mapping field. Extension to an arbitrary system ($N$) size and/or dimensions ($d$) is trivial due to the noninteracting nature of the oscillators in all directions. The physical potential energy for this model is given by $U\left(x_i\right)=\frac{1}{2} m\omega^2 x_i^2$, which results in the following effective potential (eq~\ref{eq:U_eff_def})
\begin{align}
\label{eq:V_hoq}
V^{\rm HOq}\left({\bf x}, \beta\right) = \frac{1}{n}\sum_{i=0}^{n-1} \frac{1}{2} m \omega_n^2 \left(x_i - x_{i+1} \right)^2 
+  \frac{1}{n} \sum_{i=0}^{n-1} \frac{1}{2}  m \omega^2 x_i^2
\end{align}

In order to derive the mapping field, we first need to decompose this Cartesian form into a decoupled representation using new coordinates. For this purpose, we adopt the recently developed HO normal mode and staging coordinates\cite{moustafa2024staging}. Using $\bf y$ as a generic symbol for both coordinates, we get
\begin{subequations}
\label{eq:HOq}
\begin{align}
\label{eq:HOq_V}
V^{\rm HOq} \left( {\bf y} , \beta\right) &=  \frac{1}{2}\sum_{k=0}^{n-1} \lambda_k y_k^2 \\
\label{eq:HOq_A}
\beta A^{\rm HOq}\left(\beta\right) &=  \frac{n}{2} \ln\left(\frac{2\pi \hbar^2 \beta}{mn}\right)
+  \frac{1}{2}\sum_{k=0}^{n-1}  \ln \left( \frac{\beta \lambda_k}{2\pi}\right) \\
\label{eq:HOq_E}
E^{\rm HOq} &= \frac{n}{2\beta}  - \sum_{k=0}^{n-1}  g_k
\end{align}
\end{subequations}
where $\lambda_k$ is the spring constant associated with the coordinate $y_k$ of mode $k$ and $g_k$ is a mode-based Grüneisen parameter, 
\begin{align}
\label{eq:g_k_def}
g_k  \equiv -\frac{1}{2\beta \lambda_k} \frac{\partial \left(\beta \lambda_k\right)}{\partial \beta}  
\end{align}
The definition of both $\lambda_k$ and $g_k$ parameters depends on the coordinate type, which we specify below for both HO NM and staging.

The mapping velocity can then be obtained by solving the mapping differential equation (eq~\ref{eq:mapping_0}) in the new coordinates, and then transforming back to the Cartesian space to get $\dot{\bf x}$. Once we have the mapping field, we can use the generalized PI estimator expressions in regular Cartesian coordinates (eq~\ref{eq:ECv_ma}). Using eq~\ref{eq:HOq}, the mapping equation in the transformed coordinates become
\begin{align}
 \frac{n}{2\beta}  - \sum_{k=0}^{n-1}  g_k = \frac{n}{2\beta} - \sum_{k=0}^{n-1} \frac{\partial \dot y_{k}}{\partial y_k} -  \sum_{k=0}^{n-1} \beta \lambda_k g_k y_{k}^2  +  \sum_{k=0}^{n-1} \beta \lambda_k y_{k} \dot y_{k}
 \end{align}
Using the fact that each mode is independent, it is simple to show that the solution is
\begin{subequations}
\label{eq:mapping_hmaq}
\begin{align}
\label{eq:ydot}
\dot y_{k} = g_k y_{k}
\end{align}
or, in matrix notation
\begin{align}
\label{eq:ydot_vec}
\dot{\bf y} = G {\bf y}    
\end{align}
\end{subequations}
where $\bf y$ is an $n\times 1$ vector of NM coordinates and $G$ is a diagonal matrix with $g_k$ components. Transforming back to Cartesian coordinates depends on whether we have HO normal mode or staging transformation, as we detail in the next section.

Once we have $\dot{\bf x}_i$ and $\ddot{\bf x}_i$ mapping fields, we can use either the Lagrangian (eq~\ref{eq:ECv_Lag_ref}), or the Eulerian (eq~\ref{eq:ECv_ma}) version of the new estimators, replacing the reference term by the harmonic contribution. For example, in the Eulerian version, we get
\begin{subequations}
\label{eq:ECv_hmaq}
\begin{align}
\label{eq:E_hmaq}
E &= E^{\rm HOq} + \left< U-K
- \beta \sum_{i=0}^{n-1} {\bf F}_i \cdot  \dot {\bf x}_i \right>   \\
\label{eq:Cv_hmaq}
\frac{ C_{\rm V}}{k_{\rm B} \beta^2}  
&= \frac{ C^{\rm HOq}_{\rm V}}{k_{\rm B} \beta^2}  + \left<-\frac{2}{\beta} K + \beta \sum_{i=0}^{n-1} {\bf F}_i  \cdot {\ddot {\bf x}_i} 
 +  2  \sum_{i=0}^{n-1}  \left( {\bf F}^{\rm phy}_i - {\bf F}^{\rm kin}_i \right)  \cdot \dot {\bf x}_i \nonumber \right. \\
&- \left.   \beta  \sum_{i=0}^{n-1}   m \omega_n^2 \left(\dot {\bf x}_i  - \dot {\bf x}_{i+1} \right)^2
- \beta \sum_{i=0}^{n-1}    \dot {\bf x}_i \cdot  H_{ii}^{\rm phy} \cdot \dot {\bf x}_i \right>  + {\rm var}\left(\hat E\right)
\end{align}
\end{subequations}
where $E^{\rm HOq}$ is obtained from eq~\ref{eq:HOq_E}, which can be differentiated to get $C_{\rm V}^{\rm HOq}$. However, a closed form expression is also available,\cite{schweizer1981,chin2023analytical,moustafa2024staging} 
\begin{align}
\label{eq:En_hoq}
E^{\rm HOq} = dN \frac{\hbar\omega }{2}\frac{\coth\left(\frac{n\alpha}{2}\right)}{\sqrt{1+\frac{1}{4}\epsilon^2}}
\end{align}
where $\alpha \equiv 2\sinh^{-1}\left(\frac{\epsilon}{2}\right)$ and $\epsilon\equiv\frac{\omega}{\omega_n}=\frac{\beta\hbar\omega}{n}$, from which the reference heat capacity can be obtained. Moreover, to reduce the finite Trotter number effects, the continuum limit ($n\to\infty$) expression could be used instead, $E^{\rm HOq}_{\infty} = dN \frac{\hbar\omega }{2} \coth\left(\frac{\beta\hbar\omega}{2}\right)$.

\subsection{Mapping Field for the HMAq Estimator}
\label{sec:hmaq_all}
In the next two sections we specify the mapping fields associated with the HMAq estimator, both in normal mode (HMAq-NM) and HO staging (HMAq-SG) coordinates.

\subsubsection{HMAq: HO Normal Mode Coordinates}
\label{sec:HMAq-NM_est}
In the HO NM coordinates case, $\lambda_k$ in eq~\ref{eq:HOq} are the eigenvalues of the Hessian ($n\times n$) matrix, which is given as the second derivatives of  $V^{\rm HOq}$ (eq~\ref{eq:V_hoq}) with respect to Cartesian coordinates. These eigenvalues are denoted here as $\Lambda_k$ and given by\cite{moustafa2024staging}
\begin{align}
\label{eq:V_HO_nm_lambda}
\Lambda_k  &= m \omega^2 + 4m \omega_n^2 \sin^2\left(\frac{\pi k}{n}\right), k=0,1,\dots,n-1
\end{align}
The normal mode coordinates vector $\bf q$ is related to the Cartesian coordinates vector $\bf x$ (both of length $n$) through a linear transformation,
\begin{align}
\label{eq:x2q_trans}
{\bf x} = \sqrt{n} A {\bf q}
\end{align}
where $A$ is a $n\times n$ matrix of the orthonormal eigenvectors associated with the Hessian matrix, hence the inverse is equal to the transpose, $A^{-1}=A^{\rm t}$. Since $A$ is independent of temperature, the Cartesian-space mapping velocity and acceleration are given by $\dot{\bf x} = A  \dot{\bf q}$ and $\ddot{\bf x} = A  \ddot{\bf q}$, respectively. In order to represent these in terms of Cartesian coordinates, we use eqs.~\ref{eq:ydot_vec} and~\ref{eq:x2q_trans},
\begin{subequations}
\label{eq:mapping_nm}
\begin{align}
\dot{\bf x} &= M^{\beta} {\bf x} \\
\ddot{\bf x} &= M^{\beta\beta} {\bf x} 
\end{align}    
\end{subequations}
where $M^{\beta}\equiv AGA^t$ and $M^{\beta\beta} \equiv A \left(\frac{\partial G}{\partial \beta}+G^2\right) A^t$ are the Cartesian-space mapping matrices for the velocity and acceleration, respectively. Using eqs.~\ref{eq:g_k_def} and~\ref{eq:V_HO_nm_lambda}, the mapping (diagonal) matrix in the NM space is,
\begin{align}
\label{eq:Gkk_nm}
G_{k,k}  \equiv -\frac{1}{2\beta \Lambda_k} \frac{\partial \left(\beta \Lambda_k\right)}{\partial \beta}  
= \frac{1}{2\beta} \; \frac{\sin^2\left(\frac{\pi k}{n}\right) -  \left(\frac{\epsilon}{2}\right)^2 }{ \sin^2\left(\frac{\pi k}{n}\right) +  \left(\frac{\epsilon}{2}\right)^2 }
\end{align}

It is interesting to note that this mapping reduces to the CVir case when using the RP kinetic energy (first term of eq~\ref{eq:V_hoq}) as a reference system for mapping. In this case, $\omega=0$ (or, $\epsilon=0$) in eq~\ref{eq:Gkk_nm} for $k>0$, which yields $G_{k,k}=\frac{1}{2\beta}$. As for the centroid mode ($k=0$), we have $\Lambda_0=0$ (eq~\ref{eq:V_HO_nm_lambda}) and hence $G_{0,0}=0$. Accordingly, $M^{\beta}=\frac{1}{2\beta}\left(I-\frac{1}{n}J\right)$, where $I$ and $J$ are the identity and ones matrices, respectively, both of size $n\times n$. Hence, using eq~\ref{eq:mapping_nm}, we get back the CVir mapping fields (eq~\ref{eq:mapping_cv}).

\subsubsection{HMAq: HO Staging Coordinates}
\label{sec:HMAq-SG}
In the HO staging coordinates $u_i$, the effective potential of the HOq is given as,\cite{moustafa2024staging}
\begin{subequations}
\begin{align}
\label{eq:V_ho_sg}
 V\left({\bf u},\beta\right) &= \sum_{i=0}^{n-1} \frac{1}{2} k_i u_i^2,\\
 \label{eq:V_ho_sg_ki}
k_i &= \frac{m\omega_n^2}{n}\left\{
\begin{array}{ll}
      2 \sinh\left(\alpha\right) \tanh\left(\frac{n\alpha}{2}\right), & i=0 \\
       \frac{\sinh\left(\left(n-i+1\right)\alpha\right)}{\sinh\left(\left(n-i\right)\alpha\right)}, & i>0,
\end{array} 
\right. \\
\label{eq:xi_ui_trans}
x_i &= \left\{
\begin{array}{ll}
u_0, \; i=0 \\
u_i + \frac{\sinh\left(\alpha\right) u_0 + \sinh\left(\left(n-i\right)\alpha \right) x_{i-1} }{\sinh\left(\left(n-i+1\right) \alpha\right)}, \; i>0
\end{array} 
\right.
\end{align}
\end{subequations}
where $\alpha$ is defined earlier (Section~\ref{sec:hmaq_estimator}). Note that the inverse transformation (${\bf x}\leftarrow {\bf u}$) is recursive (from $i=0$ to $n-1$). Moreover, it scales linearly with the Trotter number, unlike the quadratic scaling of normal modes. Mathematically, this transformation can also be expressed in a matrix form, similar to NM (eq~\ref{eq:x2q_trans}), with the exception that the transformation matrix is a lower triangular one. However, this would yield a quadratic scaling with $n$, or $n\log\left(n\right)$ using discrete Fourier transform.

Application of eq~\ref{eq:ydot} to HO staging coordinates yields
\begin{subequations}
\label{eq:mapping_sg_sol}
\begin{align}
\label{eq:udot_sol}
\dot{u}_i &= \gamma_i u_i \\
\gamma_i &\equiv   - \frac{1}{2\beta k_i} \frac{\partial \beta k_i}{\partial \beta}
\end{align}
\end{subequations}
where $\gamma_i$ parameters, and their derivative, are given in Section S2 of the Supporting Information. The associated mapping velocity in the Cartesian coordinates is also given in a staging fashion. Using eqs.~\ref{eq:xi_ui_trans} and~\ref{eq:udot_sol} yields the following recursive scheme (from $i=0$ to $n-1$),
\begin{subequations}
\label{eq:mapping_sg}
\begin{align}
\dot x_i  = \left\{
\begin{array}{ll}
\gamma_0  x_0, \; i=0\\
\gamma_i \left(  x_i  - {A}_i  x_0 - {B}_i  x_{i-1}  \right)  \\
+\dot{A}_i  x_0 +\dot{B}_i  x_{i-1}+{A}_i \dot x_0 +  {B}_i \dot x_{i-1},\;\; i=1,2,\dots,n-1
\end{array} 
\right.    
\end{align}
The associated acceleration is then obtained through differentiation,
\begin{align}
\ddot x_i = \left\{
\begin{array}{ll}
\left( \dot{\gamma}_0 + \gamma_0^2 \right)  x_0,  \\
\left({\dot \gamma}_i + \gamma^2_i\right)  \left(   x_i - A_i  x_0 -  B_i  x_{i-1} \right)
+ {\ddot A}_i  x_0
+ {\ddot B}_i   x_{i-1}  \\
+ 2 {\dot A}_i \dot x_0 + 2 {\dot B}_i \dot x_{i-1} + A_i   \ddot x_0 + B_i  \Delta \ddot x_{i-1}\;\; i=1,2,\dots,n-1 
\end{array} 
\right.    
\end{align}
\end{subequations}
The $A_i$ and $B_i$ parameters, and their derivatives with respect to $\beta$, are given in Section S2 of the Supporting information. Similar to the staging coordinates transformation, these recursive schemes could also be written using a lower triangular matrix form; however, we avoid this approach due to the aforementioned reasons. 

\subsubsection*{Averaging over multiple starting beads}
\begin{figure}
\includegraphics[width=0.44\textwidth]{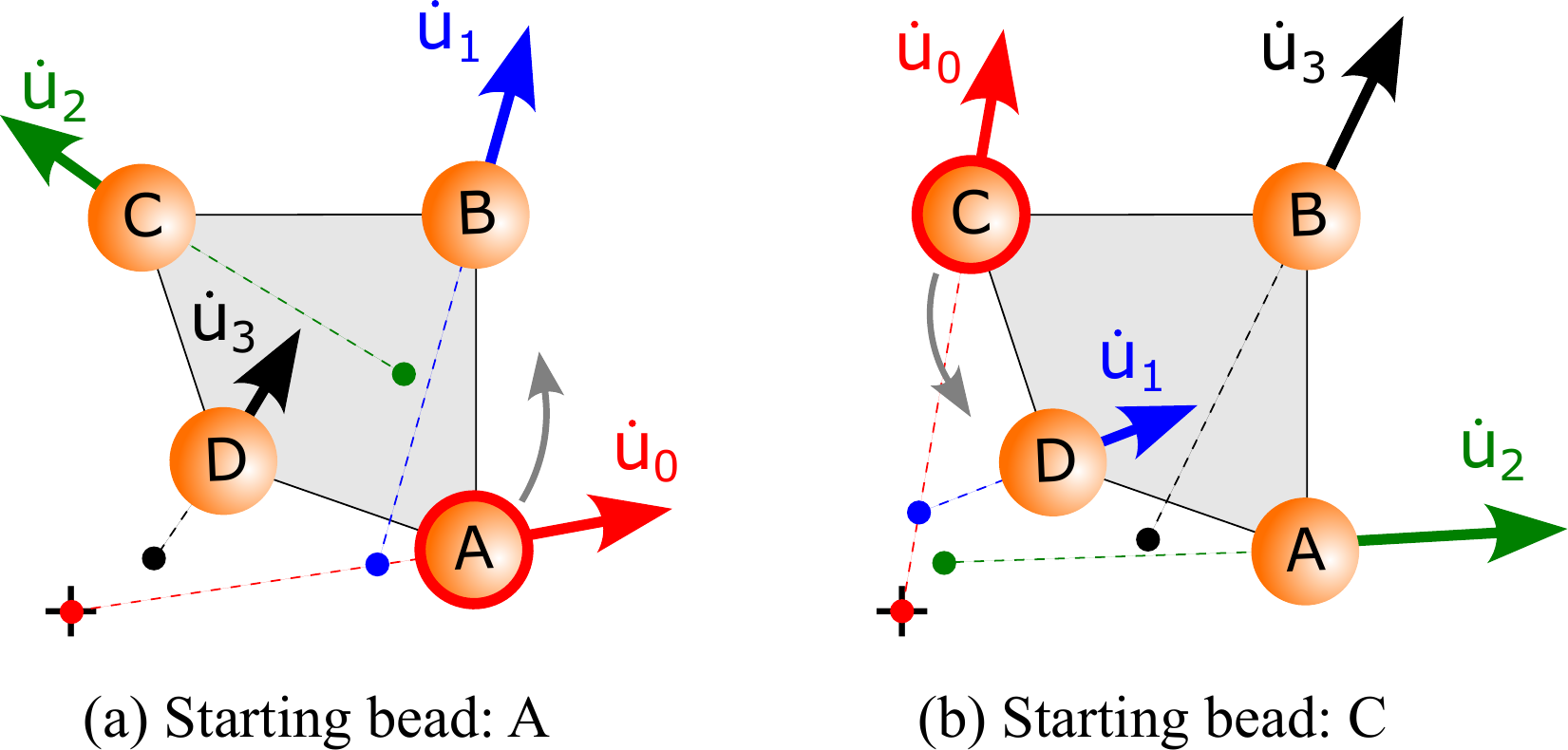}
\caption{\label{fig:starts} Illustrative example of the HMAq-SG estimator with two starts from the same beads conformation: bead ``A'' (left) and bead ``C'' (right). The dashed lines represent the HO staging coordinates, while the arrows are the associated mapping velocities. The HMAq-SG estimates produced from these mapping fields are then averaged to obtain a more precise results (see Section~\ref{sec:HMAq-SG}). }
\centering
\end{figure}

Unlike all other estimators, the staging mapping is built starting from some arbitrary bead (labeled ``0''). However, both the value and precision of ensemble averages do not depend on this choice.  With this freedom, the statistical fluctuations can be further reduced by averaging over mappings from different starting points for a given configuration. Figure~\ref{fig:starts} shows a simplified example for a ring-polymer of four beads, with two starting beads. This additional averaging is applied to terms involving mapping fields in the mapped averaging expressions (eq~\ref{eq:ECv_ma}),
\begin{subequations}
\label{eq:ECv_nmsg_starts}
\begin{align}
\label{eq:E_nmsg_starts}
E &= E^{\rm HOq} + \Big\langle U-K\Big\rangle 
-  \llangle[\Big]\beta \sum_{i=0}^{n-1} {\bf F}_i \cdot  \dot {\bf x}_i \rrangle[\Big] \\
\label{eq:Cv_nmsg_starts}
\frac{ C_{\rm V}}{k_{\rm B} \beta^2}  
&= \frac{ C^{\rm HOq}_{\rm V}}{k_{\rm B} \beta^2}  - \Big\langle\frac{2}{\beta} K\Big\rangle +\llangle[\Big] \beta \sum_{i=0}^{n-1} {\bf F}_i  \cdot {\ddot {\bf x}_i} 
 +  2  \sum_{i=0}^{n-1}  \left( {\bf F}^{\rm phy}_i - {\bf F}^{\rm kin}_i \right)  \cdot \dot {\bf x}_i   \nonumber \\
&- \beta  \sum_{i=0}^{n-1}   m \omega_n^2 \left(\dot {\bf x}_i  - \dot {\bf x}_{i+1} \right)^2
- \beta \sum_{i=0}^{n-1}    \dot {\bf x}_i \cdot  H_{ii}^{\rm phy} \cdot \dot {\bf x}_i \rrangle[\Big] + {\rm var}\left(\hat E\right)
\end{align}
\end{subequations}
where the double angle brackets indicate the total average over all the samples
\begin{align}
\label{eq:avg_starts}
\llangle \hat X \rrangle  \equiv  \frac{1}{n_{\rm c} n_{\rm s}} \sum_{i=1}^{n_{\rm c}}  \sum_{j=1}^{n_{\rm s}} \hat X_{ij}
\end{align}
where $n_{\rm c}$ is the number of configurations, $n_{\rm s}$ is the number of starting beads used for each configuration, and $\hat X_{ij}$ represents a sample associated with a configuration $i$ and starting bead $j$. Therefore, we get $n_{\rm s}\times$ more samples compared to using only one starting bead, which results in higher precision. Notice that the ${\rm var}(\hat E)$ term represents the variance of these $n_{\rm c}n_{\rm s}$ extended raw samples, i.e., ${\rm var}(\hat E)=\llangle\hat E^2\rrangle - \llangle\hat E\rrangle^2$

Although the choice of these starting beads is arbitrary, we select them uniformly (i.e., equal separation along the ring-polymer) to ensure more independent samples. This, however, requires the number of beads to follow this binary form: $n=2^p$, with $p$ being an integer. In this case, the number of starting points are also multiples of $2$. For instance, if $n=4$ ($p=2$), then the number of starting points could be $1$ (one start), $2$ (average over two starts; beads $0$ and $3$), or $4$ (average over beads $0, 1, 2, 3$). Figure~\ref{fig:starts} shows this example, using the case of two starting beads, bead ``A'' (left) and bead ``C'' (right).

\begin{table*}[]
\caption{\label{tab:estimators}Path integral expressions for the total energy $E$ and isochoric heat capacity $C_{\rm V}$, using standard (thermodynamic and centroid virial) and harmonically mapped averaging (HMAc and HMAq) estimators. The HMAq estimators can be implemented in either HO normal mode (HMAq-NM) or HO staging (HMAq-SG) coordinates (see Section~\ref{sec:hmaq_all}). The expressions are explicit evaluation of the Eulerian representation (eq~\ref{eq:ECv_ma}), which rely on forces and Hessian matrix being available. When this is not the case, the Lagrangian version (eq~\ref{eq:ECv_Lag_ref}) can be employed, with total derivatives evaluated using finite difference schemes (see Section S1 of the Supporting Information). For application of HMA estimators to unbound models, such as LJ, a center of mass contribution need to be subtracted from the energy ($\frac{d}{2\beta}$) and heat capacity ($\frac{C_{\rm V}}{k_{\rm B}\beta^2}$).
}
\begin{tabular}{|cc|ccc|c|}
\hline
\multicolumn{2}{|c|}{\multirow{2}{*}{Estimator}}                                                                                 & \multicolumn{3}{c|}{Mapping}  & \multirow{2}{*}{Expression}  \\ \cline{3-5}
\multicolumn{2}{|c|}{}                                                                                                           & \multicolumn{1}{c|}{reference} & \multicolumn{1}{c|}{$\dot{\bf x}_i$} & $\ddot{\bf x}_i$ &      \\ \hline
\multicolumn{1}{|c|}{\multirow{2}{*}{\rotatebox{90}{Standard\;\;\;\;\;}}}                 
& TD                                                 & \multicolumn{1}{c|}{\begin{tabular}[c]{@{}c@{}} classical IG
\\ $E^{\rm IGc} =\frac{dNn}{2\beta}$ \;\;\;\\ $C_{\rm V}^{\rm IGc}=\frac{dNn}{2} k_{\rm B}$
\end{tabular}} & \multicolumn{1}{c|}{0}  & 0   & 
$
\begin{array} {lll} 
E &=& E^{\rm IGc} +  \left< U-K  \right>
\\
\frac{ C_{\rm V}}{k_{\rm B}\beta^2}  &=& 
\frac{ C_{\rm V}^{\rm IGc}}{k_{\rm B}\beta^2}  - \frac{2}{\beta} \left< K  \right> + {\rm var}\left(\hat E\right)
\end{array}
$ 
\\ 
\cline{2-6} 
\multicolumn{1}{|c|}{}                                               
& 
\begin{tabular}[c]{@{}l@{}}CVir\end{tabular} & \multicolumn{1}{c|} {\begin{tabular}[c]{@{}c@{}} quantum IG
\\ $E^{\rm IGq} =\frac{dN}{2\beta}$ \;\;\; \\ $C_{\rm V}^{\rm IGq}=\frac{dN}{2} k_{\rm B}$
\end{tabular}} 
& \multicolumn{1}{c|}{$\frac{{\bf x}_i - {\bf x}_{\rm c}}{2\beta} $}     &       $-\frac{{\bf x}_i - {\bf x}_{\rm c}}{4\beta^2}$ &
$
\begin{array}{lll}
E &=& E^{\rm IGq}  + \left<  U   - \frac{1}{2}  \sum\limits_{i=0}^{n-1}  {\bf F}_i^{\rm phy} \cdot   \left( {\bf x}_i - {\bf x}_{\rm c}\right)  \right> 
\\
\frac{C_{\rm V}}{k_{\rm B} \beta^2}   
&=& \frac{ C_{\rm V}^{\rm IGq}}{k_{\rm B}\beta^2} + \frac{1}{4\beta} \left< 3  \sum\limits_{i=0}^{n-1}  {\bf F}_i^{\rm phy}  \cdot  \left( {\bf x}_i - {\bf x}_{\rm c}\right)  \right> \\
&-&  \frac{1}{4\beta} \left< \sum\limits_{i=0}^{n-1} \left( {\bf x}_i - {\bf x}_{\rm c}\right) \cdot  H_{ii}^{\rm phy} \cdot \left( {\bf x}_i - {\bf x}_{\rm c}\right) \right> +  {\rm var}\left(\hat E\right)
\end{array}
$
\\ \hline
\multicolumn{1}{|c|}{\multirow{2}{*}[6ex]{\rotatebox{90}{Harmonically mapped averaging}}} & HMAc                                                      & \multicolumn{1}{c|}{\begin{tabular}[c]{@{}c@{}} classical HO
\\ $E^{\rm HOc} =\frac{dN}{\beta}$ \;\;\; \\ $C_{\rm V}^{\rm HOc}=dN k_{\rm B}$
\end{tabular}}       
& \multicolumn{1}{c|}{$\frac{{\bf x}_i - 2{\bf x}_{\rm c}}{2\beta}$}     &       $-\frac{{\bf x}_i - 2{\bf x}_{\rm c}}{4\beta^2}$  &
$
\begin{array}{lll}
E   &=&  E^{\rm HOc} + \left< U  
- \frac{1}{2}\sum\limits_{i=0}^{n-1}  {\bf F}_i^{\rm phy}  \cdot \left(  {\bf x}_i - 2 {\bf x}_{\rm c} \right)  \right>   \\
\frac{ C_{\rm V}}{k_{\rm B} \beta^2}    &=& 
\frac{ C_{\rm V}^{\rm HOc}}{k_{\rm B}\beta^2} + \frac{1}{4\beta}\left< 3  \sum\limits_{i=0}^{n-1}  {\bf F}_i^{\rm phy}  \cdot  \left( {\bf x}_i - 2{\bf x}_{\rm c}\right) + 2 \sum\limits_{i=0}^{n-1}  {\bf F}_i^{\rm phy}  \cdot  {\bf x}_{\rm c} \right>  \\
&-& \frac{1}{4\beta}\left<   \sum\limits_{i=0}^{n-1} \left( {\bf x}_i -2 {\bf x}_{\rm c}\right) \cdot  H_{ii}^{\rm phy} \cdot \left( {\bf x}_i - 2{\bf x}_{\rm c}\right)
\right>  + {\rm var}\left(\hat E\right)
\end{array}
$
\\
\cline{2-6} 
\multicolumn{1}{|c|}{}                                               & HMAq                                                      & \multicolumn{1}{c|}{\begin{tabular}[c]{@{}c@{}} quantum HO
\\ $E^{\rm HOq} = \frac{dN\hbar\omega }{2}\frac{\coth\left(\frac{n\alpha}{2}\right)}{\sqrt{1+\frac{1}{4}\epsilon^2}}$\\ $C_{\rm V}^{\rm HOq} =\frac{\partial E^{\rm HOq}}{\partial T}$ \;\;\;\;\;\;\;\;\;\;
\end{tabular}}       
& \multicolumn{2}{c|}{ \multirow{2}{*}{\begin{tabular}[c]{@{}l@{}}HMAq-NM: eq~\ref{eq:mapping_nm} \\ HMAq-SG: eq~\ref{eq:mapping_sg} \end{tabular}}}  &
$
\begin{array}{lll}
E   &=&
E^{\rm HOq} + \left< U-K
- \beta {\bf F}^{\rm eff}_i  \cdot  \dot {\bf x}_i \right> \\
\frac{ C_{\rm V}}{k_{\rm B} \beta^2}  
&=&  \frac{C_{\rm V}^{\rm HOq}}{k_{\rm B}\beta^2}
+ {\rm var}\left(\hat E\right)\\
&+& \left<
 \beta \sum\limits_{i=0}^{n-1} {\bf F}^{\rm eff}_i  \cdot {\ddot {\bf x}_i} -\frac{2}{\beta} K + 2  \sum\limits_{i=0}^{n-1}  \left( {\bf F}_i - {\bf F}^{\rm kin}_i \right)  \cdot \dot {\bf x}_i \right> \\
&-& \beta \left<    \sum\limits_{i=0}^{n-1}   m \omega_n^2 \left(\dot {\bf x}_i  - \dot {\bf x}_{i+1} \right)^2
+ \sum\limits_{i=0}^{n-1}    \dot {\bf x}_i \cdot  H_{ii} \cdot \dot {\bf x}_i \right>
\end{array}
$
\\ \hline
\end{tabular}
\end{table*}

\subsection{Computational Details} 
We consider two models, a one-dimensional anharmonic oscillator (AO) and a three-dimensional Lennard-Jones (LJ) FCC crystal. We adopt an anisotropic AO model, with cubic and quartic terms, 
\begin{align}
\label{eq:U_ao_1d}
U_{\rm AO}\left(x\right) &=\frac{1}{2}m\omega^2 x^2 + k_3 x^3 + k_4 x^4,
\end{align}
where we use $m=\omega=1$ and $k_3=k_4=0.1$ throughout this work.

For the FCC crystal, we use the standard 12-6 Lennard-Jones (LJ) model, truncated at $r_{\rm c}=3.0$,
\begin{align}
U_{\rm LJ}\left(r\right) =  4 \epsilon_{\rm LJ} \left[ \left(\frac{\sigma_{\rm LJ}}{r}\right)^{12} - \left(\frac{\sigma_{\rm LJ}}{r}\right)^{6}\right],
\end{align}
where $r$ is the pair separation. The energy and distance in this model are given in the reduced units of the LJ size ($\sigma_{\rm LJ}$) and energy ($\epsilon_{\rm LJ}$) parameters, which are set to one. The FCC crystal is made of  $4\times4\times4$ supercell ($N=256$), with a number density of $\rho=1.0$. However, for the finite size effects analysis, we consider multiple systems sizes ($N=108, 256, 500, 864, 1372$). Unlike the AO case, the $\omega$ frequency used with the HMAq EC reference is not given explicitly in the LJ model. However, as mentioned earlier, this is merely an estimator parameter, which could be (in principle) chosen to optimize performance (i.e., minimize uncertainty). We choose to define $\omega$ of the Einstein crystal reference based on the self term of the force constant (Hessian) matrix $\Phi_{ii}=\frac{\partial U_{\rm LJ}}{\partial x_i \partial x_i}=218.22018$, such that $\omega=\sqrt{\frac{\Phi_{ii}}{m}}$, where $m=1$ is the atomic mass. Performance of the HMAq estimators did not show sensitivity to other values in the vicinity of this choice.  

We use path integral molecular dynamics (PIMD) simulation to sample configurations for both AO and LJ models. However, relative performance should not depend on the sampling method. The simulations are carried out in the NVT ensemble, with the temperature controlled using white noise Langevin (stochastic) thermostat, with a BAOAB splitting order as described elsewhere.\cite{moustafa2024staging} The thermostat friction coefficient was set to $\gamma=\omega$, which provides results not far from optimum\cite{liu2016simple}. For its efficient sampling, we ran simulations in the recently introduced HO staging coordinates.\cite{moustafa2024staging} The method uses $\omega$ as an input, for which we use the aforementioned values for the AO and LJ models.

A simulation length of $10^5$ MD steps was used for both models, after $10^4$ steps of equilibration. Due to difference in oscillation frequency and convergence rates, we use different step sizes ($\Delta t=0.2$ for AO and $0.01$ for LJ). These values are based on a convergence analysis, which we present in Section S4 of the Supporting Information. The analysis also show a faster convergence of HMA estimators in comparison to the centroid virial. The number of beads used is given by $n=20\beta\hbar\omega$, which yields statistically converged results for both models, as observed earlier.\cite{moustafa2024staging,moustafa2024fse} 

We estimated the statistical uncertainties (error bars) in ensemble averages of $E$ and $C_{\rm V}$ from a single run, using the block averaging technique of $100$ blocks.\cite{book2023understanding}  While it is straightforward for the case of energy, a propagation of error is required for the case of heat capacity, as we detailed in a previous work.\cite{moustafa2024staging}  All error bars correspond to 68\% confidence limits, i.e., $\pm \sigma$.    

All the PIMD simulations were performed using the Etomica simulation package,\cite{schultz2015etomica} which can be accessed on GitHub at: https://github.com/etomica/etomica/tree/path\_integral

\section{RESULTS AND DISCUSSION}
\label{sec:results}
\begin{figure}
\includegraphics[width=0.5\textwidth]{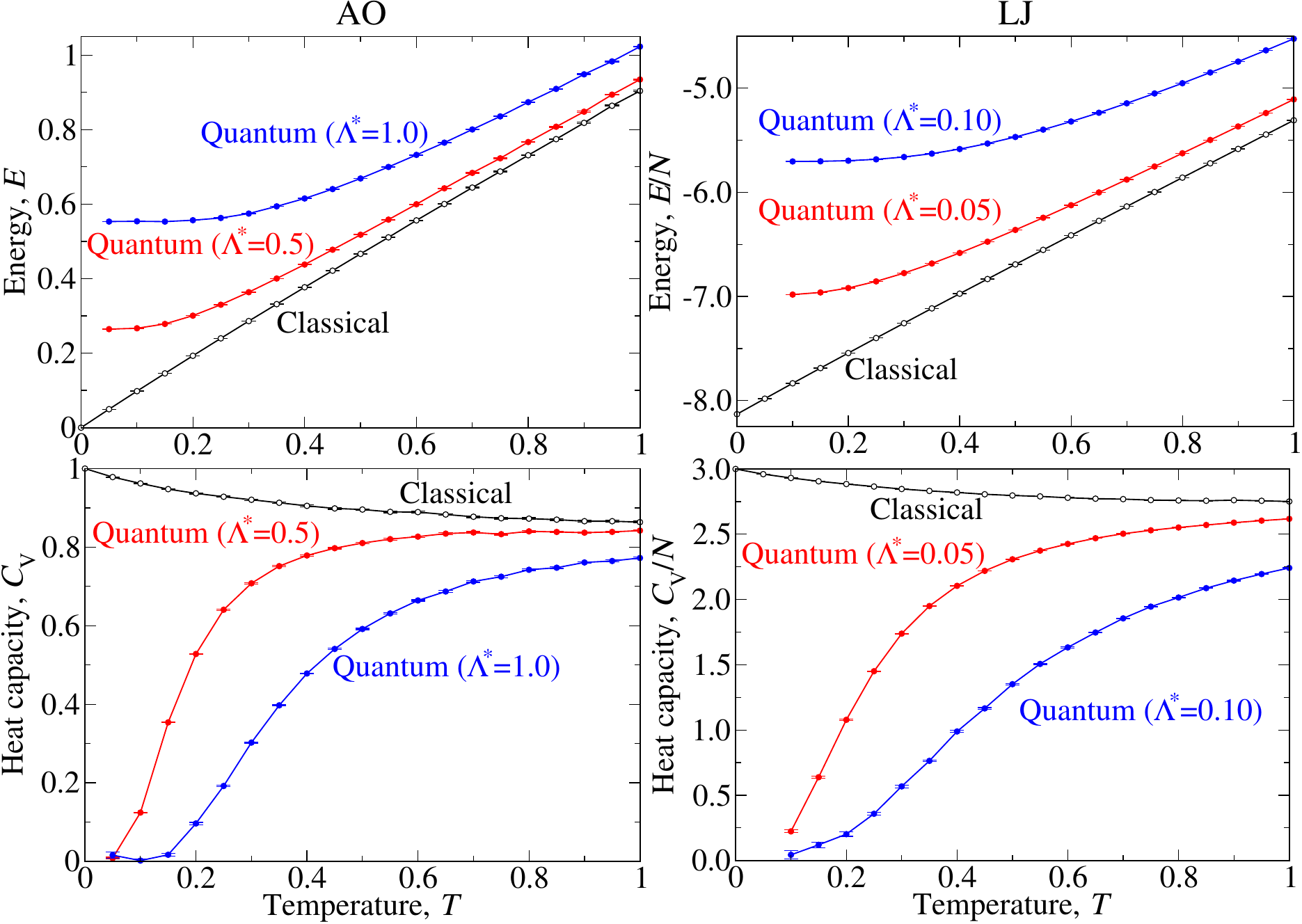}
\caption{\label{fig:ECv_T_all} Temperature dependence of the energy (top)
and heat capacity  (bottom) of the one-dimensional AO (left) and LJ (right) models. Two quantumness ($\Lambda^*$) values are considered with each model. Results from classical representation ($n=1$) are presented to show the extent of
quantumness present. Error bars, here and throughout this work, correspond to a 68\% confidence
limits. Most error bars are
smaller than the symbol size and lines join the data points
as a guide to the eye. The data were generated using the HMAq-NM estimator.}
\centering
\end{figure}

We present the performance results of the new HO-based estimators (HMAc, HMAq-NM, and HMAq-SG) in comparison to the traditional centroid virial estimator. The assessment is based on the precision and accuracy of the ensemble averages of the total energy $E$ and heat capacity $C_{\rm V}$ of the 1D AO model and the 3D LJ crystal. However, to gain an insight on the quantumness degree, we provide in Fig.~\ref{fig:ECv_T_all} a comparison of the quantum energy (top) and heat capacity (bottom) of the AO (left) and LJ (right) models against the classical values. It is evident that the intrinsic quantumness values considered ($\Lambda^*=0.5$ and $1.0$ for AO; $\Lambda^*=0.05$ and $0.1$ for LJ) results in substantial quantum effects at the entire temperature range considered. The data were generated using the HMAq-NM estimator, which is the most precise choice as we show below.

\begin{figure}
\includegraphics[width=0.5\textwidth]{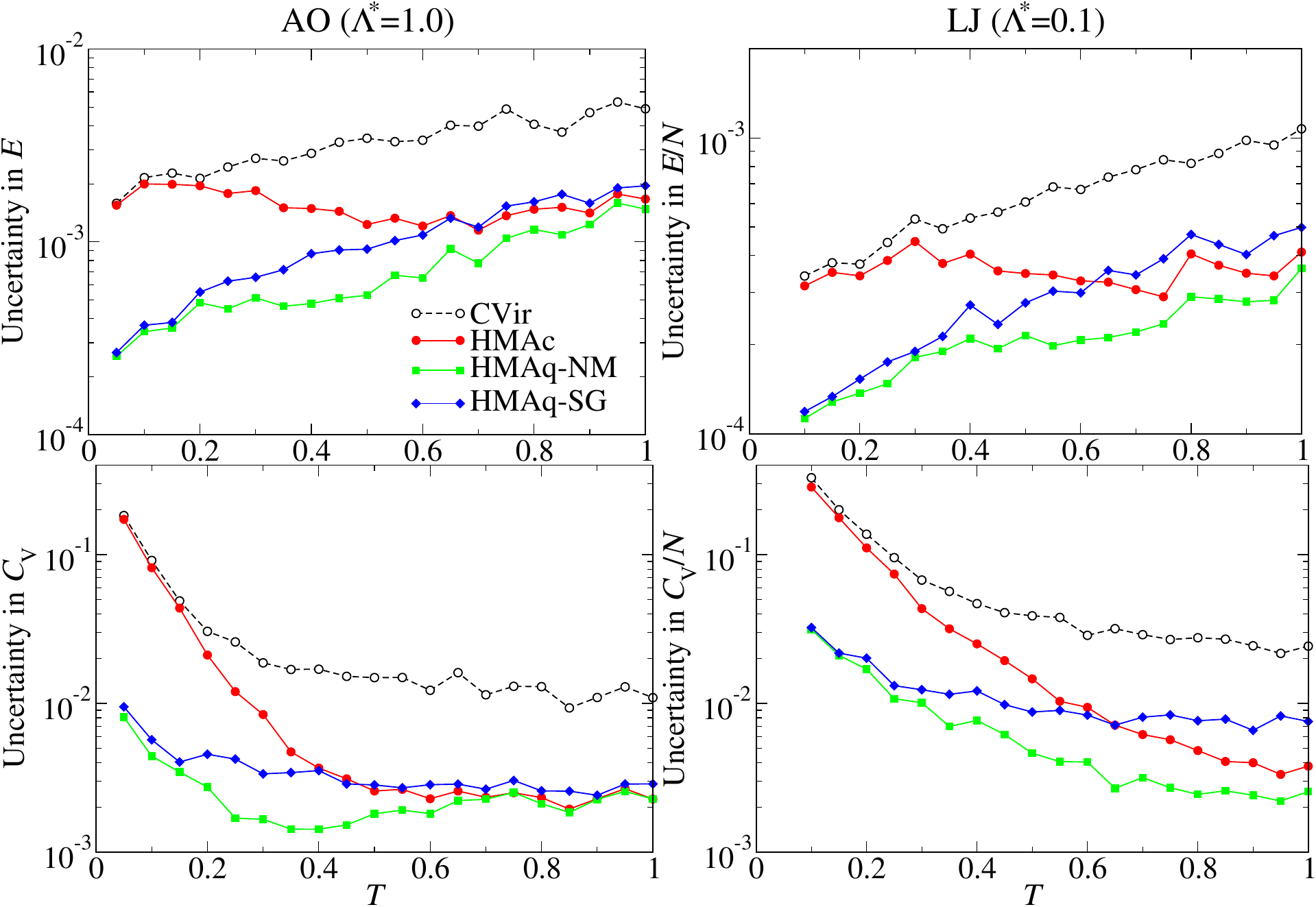}
\caption{\label{fig:err_T_all} Temperature dependence of the statistical uncertainty in ensemble averages of energy (top) and heat capacity (bottom) for the AO (left) and LJ (right) models, using the centroid virial and HMA estimators.}
\centering
\end{figure}

\subsection{Effect of Temperature on Precision}
In this section, we compare the performance of PI estimators in terms of the precision they provide for ensemble averages, at different temperatures. Figure~\ref{fig:err_T_all} shows the temperature dependence of the statistical uncertainties in the total energy (top) and heat capacity (bottom) of the AO (left) and LJ (right) models at the highest intrinsic quantumness considered ($\Lambda^*=1.0$ for AO and $0.1$ for LJ). A key observation here is that all the HMA estimators consistently exhibit higher precision than the centroid virial estimator, across the entire temperature range. However, the relative performance depends of the specific HMA ``flavor'', with the HMAq-NM method providing the most precise estimates. This is to be expected because this estimator uses a quantum HO reference for mapping, which is, relatively, the closest model that resembles the actual system. Therefore, the corresponding mapping field is able to describe how the beads scale as $\beta$ changes (see Fig.~\ref{fig:mapping}). This results in a small anharmonic contribution (eq~\ref{eq:ECv_Lag_ref}) and, hence, small fluctuations relative to other estimators. The improvement in precision provided by the HMAq-NM estimator does not arbitrarily increase at low temperatures because quantum systems do not behave purely harmonically in the $T\to 0$ limit, due to the anharmonic zero-point energy.

\begin{figure}
\includegraphics[width=0.5\textwidth]{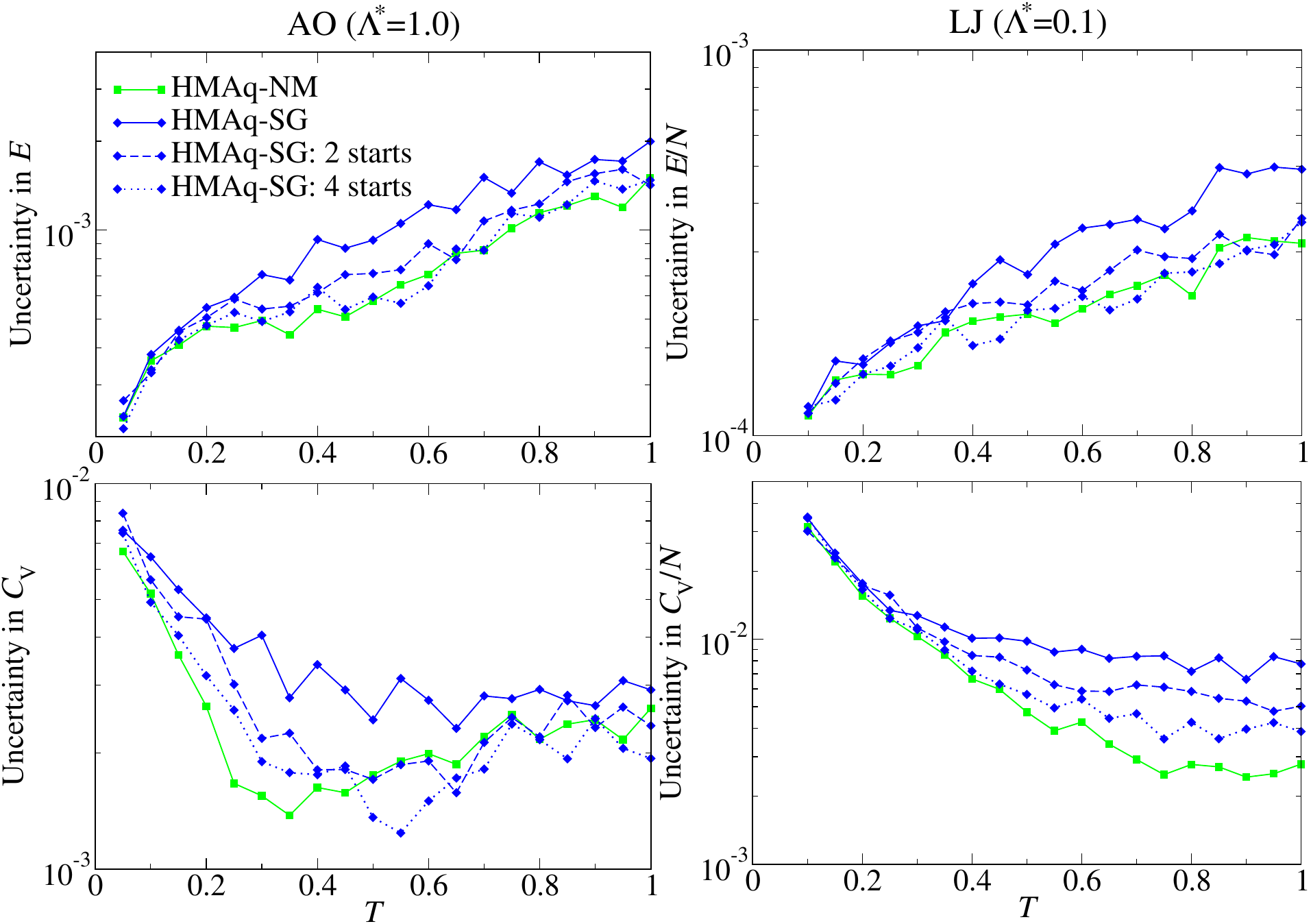}
\caption{\label{fig:errECv_T_shifts_all}Dependence of the precision in estimating the energy (top) and heat capacity (bottom) using the HMAq-SG method on the number of the starts used in the estimator. The HMAq-NM data are shows as a reference that represents the highest precision possible.}
\centering
\end{figure}

On the other hand, the HMAc estimator offers intermediate performance between the CVir and HMAq-NM methods. At low temperatures, the precision of the HMAc estimates aligns with that of CVir, while at high temperatures, it converges toward the values obtained by HMAq-NM. The behavior at low temperatures can be attributed to the structure of the HMAc mapping, which comprises two independent components: one for the ring polymer (identical to CVir, eq~\ref{eq:mapping_cv}) and a classical centroid mapping ($\dot{\bf x}_{\rm c}=-\frac{1}{2\beta}{\bf x}{\rm c}$). Additionally, since $\langle{\bf x}_c^2\rangle$ vanishes as $T \to 0$ while $\langle\left({\bf x}_i-{\bf x}_c\right)^2\rangle$ remains finite (due to zero-point energy), the centroid mapping velocity diminishes more rapidly. Consequently, in this limit, the HMAc behaves similarly to CVir, which explains the observed precision. However, as the temperature increases, the finite mapping velocity of the centroid leads their performance to differ, with the HMAc estimator providing more precise estimates. This is due to the fact that the HMAc method prescribes an explicit mapping for the centroid degree of freedom, which is absent in CVir. Indeed, as the temperature rises further, the performance of HMAc becomes comparable to that of HMAq-NM. This similarity suggests that the centroid and internal (intra-ring) modes become increasingly decoupled as temperature increases, causing the HMAq-NM approach to behave similarly to HMAc, which already assumes decoupling.

Fig.~\ref{fig:err_T_all} shows that the HMAq-SG estimator provides the closest efficiency to the HMAq-NM estimator and outperforms the HMAc method in most states. This is expected, given that they both use the same reference (HOq) for mapping.  The staging estimator does perform slightly worse at most temperatures because it treats each bead in the ring differently (see Fig.~\ref{fig:starts}). The NM mapping treats all the beads equally, which matches the true behavior and results in a better performance than the SG case.

However, as indicated earlier, the performance of the HMAq-SG estimator could be further improved by averaging over multiple starting beads (see Fig.~\ref{fig:starts} and eq~\ref{eq:ECv_nmsg_starts}). This provides more samples and, hence, better precision than just using a single starting bead as used above. Figure~\ref{fig:errECv_T_shifts_all} presents the effect of using a few starting beads on the performance of the HMAq-SG estimator. This idea is applied to the energy (top) and heat capacity (bottom) of both the AO (left) and LJ (right) models at the highest quantumness levels considered ($\Lambda^*=1.0$ for AO and $0.1$ for LJ). Using only $2$ starting beads results in a noticeable reduction in the uncertainty to a level comparable (in most cases) to the HMAq-NM estimator. Arguably, the only exception is the heat capacity of the LJ model, which appears to require averaging over more starting beads (presumably, $4$). These multiple starts, however, only introduce a small overhead CPU cost ($<40$\% compared to not using them), suggesting a cheap data collection relative to PIMD sampling. Hence, with the comparable precision to the HMAq-NM approach, the HMAq-SG with few starting beads provides a more efficient alternative that the HMAq-NM.

\begin{figure}
\includegraphics[width=0.5\textwidth]{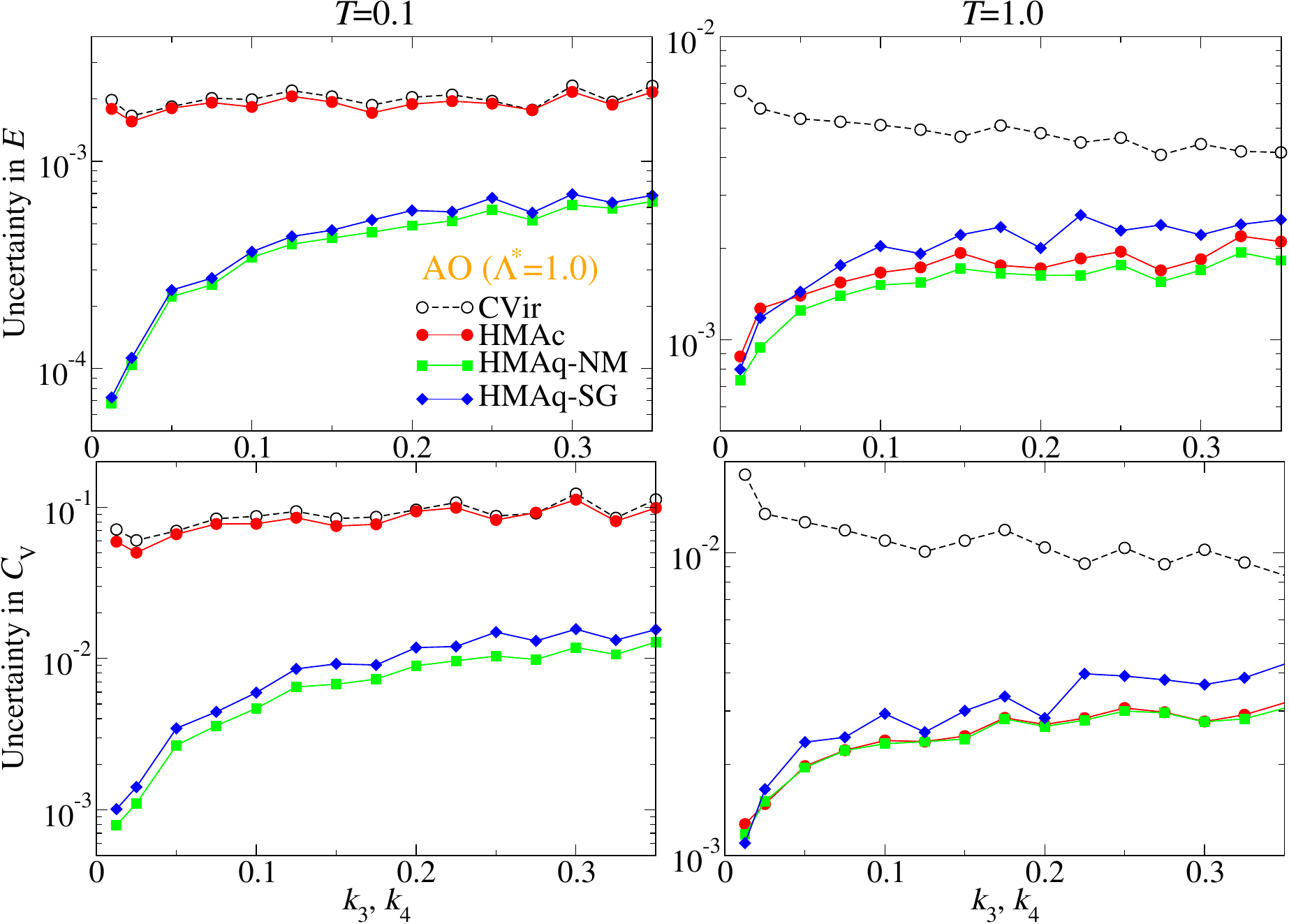}
\caption{\label{fig:errECv_k34} Effect of the anharmonicity in the AO model ($\Lambda^*=1.0$) on the precision of ensemble averages of the energy (top) and heat capacity (bottom) at temperatures $T=0.1$ (left) and $1.0$ (right).}
\centering
\end{figure}

\begin{figure*}
\centering
\subfloat[\centering AO]{{\includegraphics[width=0.48\textwidth]{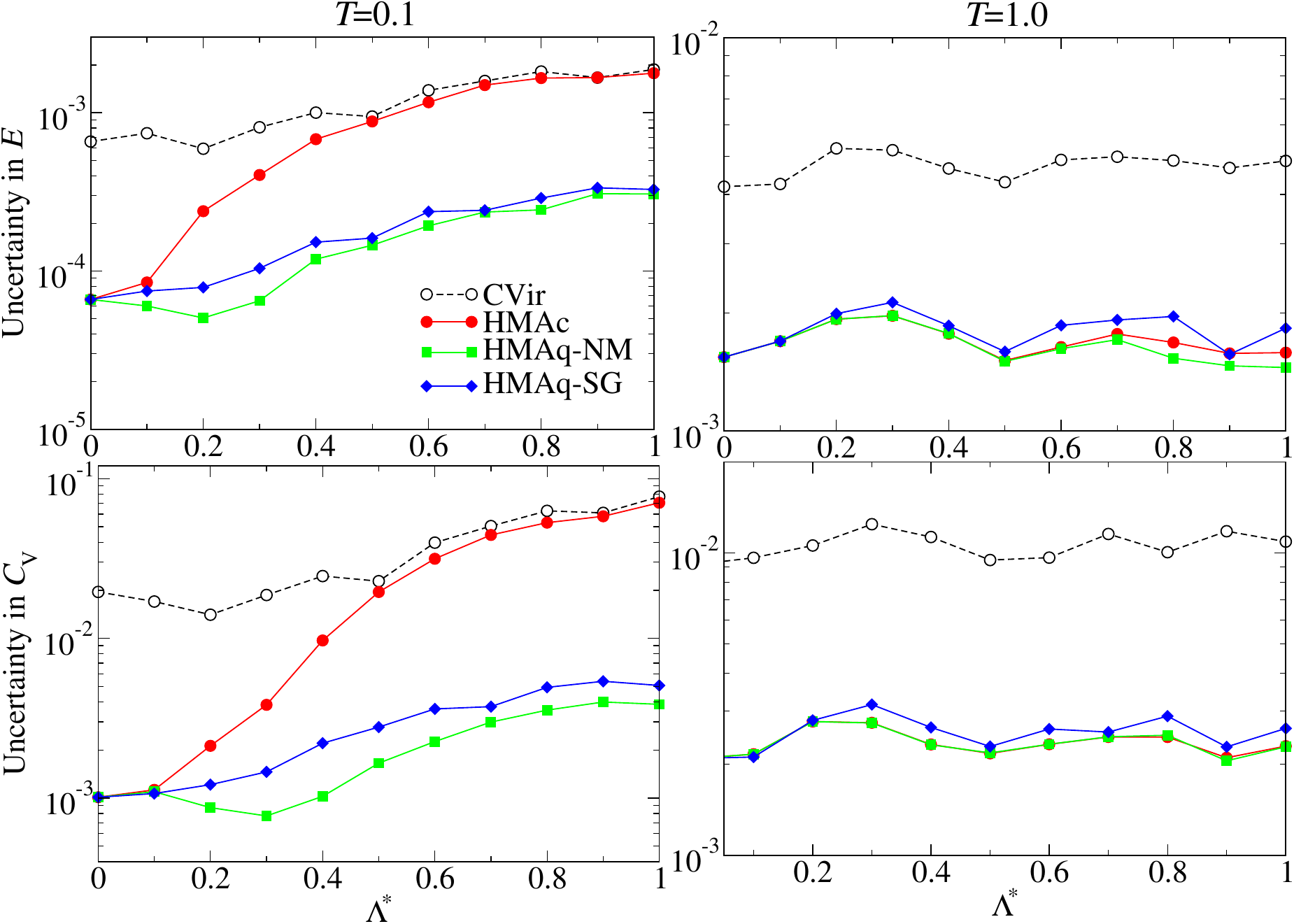} }}%
\quad
\subfloat[\centering LJ]{{\includegraphics[width=0.48\textwidth]{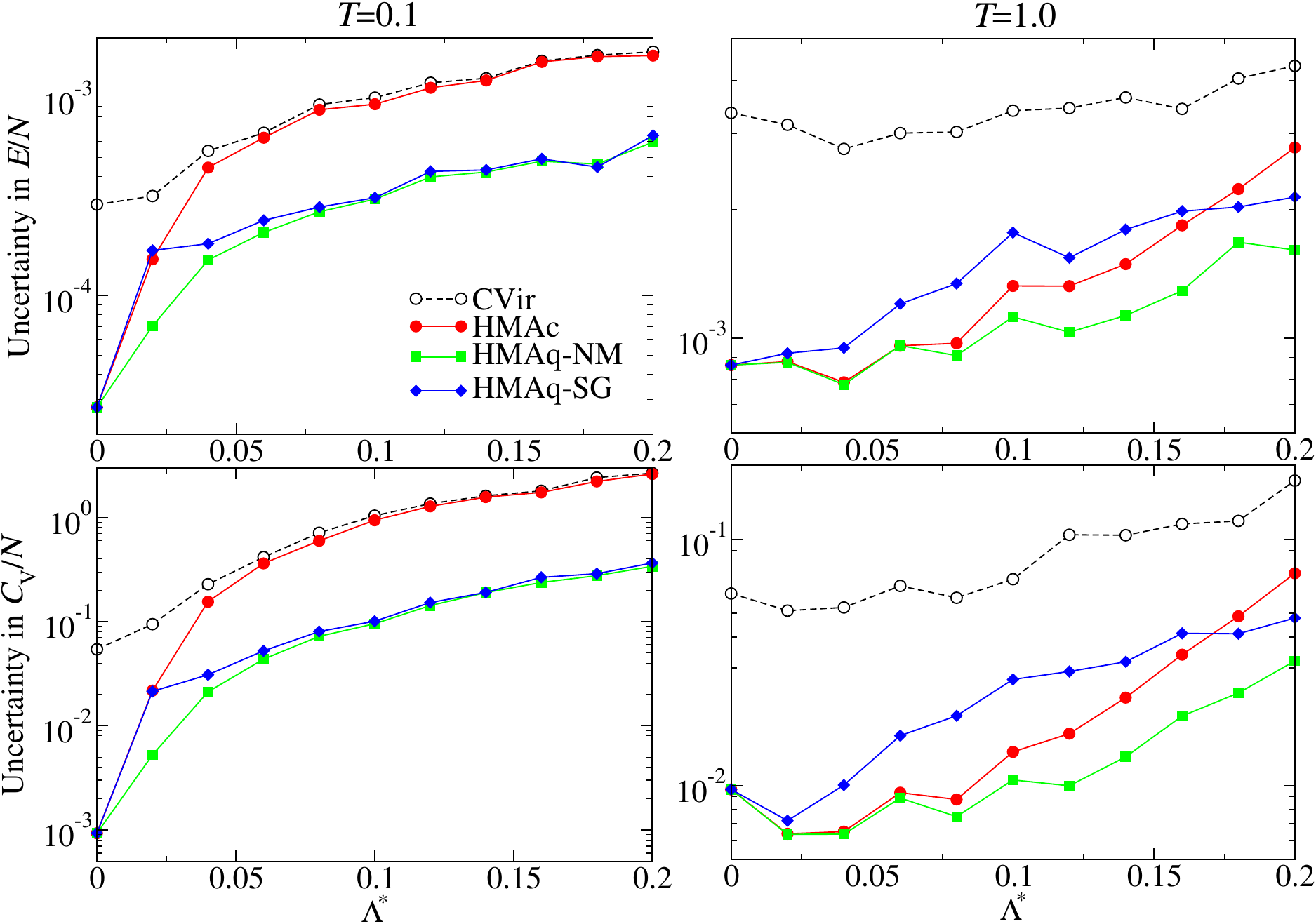} }}%
\caption{\label{fig:errECv_lambda} Dependence of the statistical uncertainty in assemble averages of the energy (top) and heat capacity (bottom) on the intrinsic quantumness ($\Lambda^*$) for the AO (left) and LJ (right) models, at low ($T=0.1$) and high ($T1.0$) temperatures. }
\end{figure*}

\subsection{Anharmonic Effects}
Since the HMA estimators are based on the HO model for the coordinates mapping, it is of interest to investigate the effect of anharmonicity on the performance. In this regard, we only consider the AO model due to its explicit anharmonic contribution (eq~\ref{eq:U_ao_1d}). Figure~\ref{fig:errECv_k34} depicts the effect of cubic ($k_3$) and quartic ($k_4$) anharmonicity, simultaneously, on the precision of estimating the energy (top) and heat capacity (bottom) at low ($T=0.1$) and high ($T=1.0$) temperatures. A first observation to report here, at both temperatures, is the significant reduction in the statistical uncertainties from the HMAq-NM and HMAq-SG estimators as the anharmonicity decreases. In fact, both estimators provide exact values (zero uncertainty) in the pure harmonic limit ($k_3=k_4=0$). This outcome is expected since the mapping is based on the harmonic behavior; hence, the precision improves as the harmonic character becomes more dominant. Nevertheless, the relative improvement compared to the CVir approach remains consistent across the entire parameter space. In contrast, the CVir estimator does not show much sensitivity to anharmonicity, apparently because it uses an IGq reference for mapping, which does not describe the HO/AO model. 
For the HMAc estimator, the previously noted similarity in its behavior to the CVir and HMAq-NM estimators at low and high temperatures, respectively, remains consistent across the entire anharmonicity domain.

\subsection{Intrinsic Quantumness Effects}
In this section, we examine the impact of intrinsic quantumness ($\Lambda^*\equiv \hbar/\sqrt{m}$) on the performance of each estimator. As mentioned earlier, the default $\Lambda^*$ values used in this study are $1.0$ (AO) and $0.1$ (LJ). Figure~\ref{fig:errECv_lambda} presents the dependence of the statistical uncertainties in the energy (top) and heat capacity (bottom) on $\Lambda^*$, for the AO (left) and LJ (right) models, both at low ($T=0.1$) and high ($T=1.0$) temperatures. The $\Lambda^*=0$ case corresponds to the classical system, which is equivalent to having a single bead ($n=1$). In this limit, the three HMA estimators become identical, which is manifested in the figure as identical uncertainties. In all cases considered, the HMAq estimators show a consistent improvement over the CVir method throughout the entire quantumness range. While this relative performance appears to be insensitive to $\Lambda^*$ for the case of AO, it shows a marginal improvement for the LJ system as the system move toward the classical limit. In contrast, the behavior of the HMAc estimator is more complex. At high temperature, it performs similarly to the HMAq estimators, again, suggesting decoupling between the centroid and internal modes. However, at low temperatures, the precision of the HMAc estimator begins to decline, approaching the performance of the CVir method as the degree of quantumness increases. This behavior is consistent with the remark made earlier that the HMAc estimator should behave like the CVir as the system becomes more quantum, either intrinsically at larger $\Lambda^*$ or thermally at lower $T$.

\begin{figure*}
\label{fig:fse_n_all}
\centering
\subfloat[\centering AO ($\Lambda^*=1.0$ and $T=0.1$)]{{\includegraphics[width=0.48\textwidth]{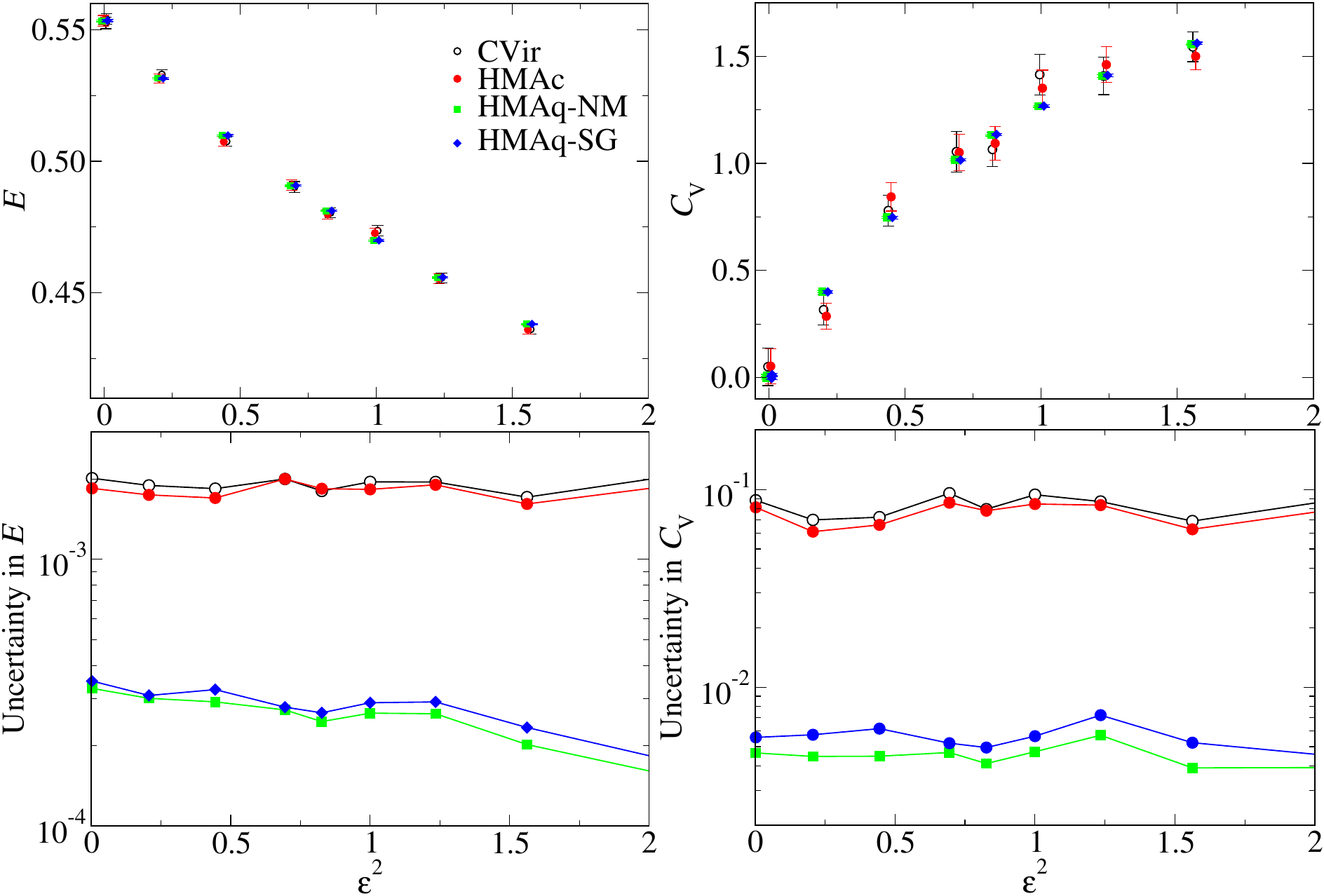} }}%
\quad
\subfloat[\centering LJ ($\Lambda^*=0.1$ and $T=0.1$)]{{\includegraphics[width=0.48\textwidth]{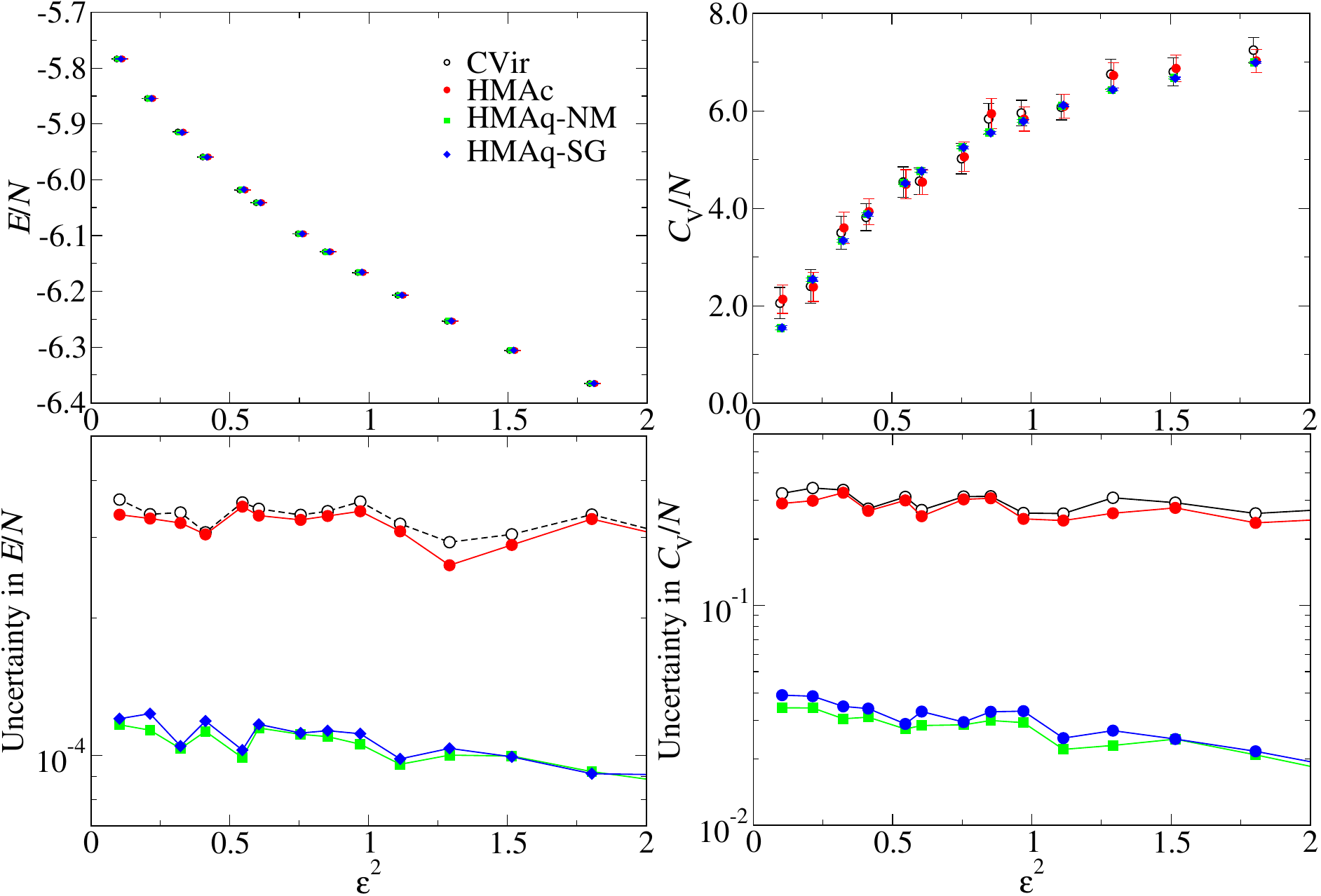} }}%
\caption{\label{fig:ECv_eps2_all} Finite Trotter number ($n$) effects on the value (top) and statistical uncertainty (bottom) of the energy and heat capacity of the AO (left) and LJ (right) models, both at $T=0.1$. The data are presented in  terms of $\epsilon^2$, where $\epsilon\equiv \beta\hbar\omega/n$. The data in the top panels are shifted slightly left/right for clarity.}
\end{figure*}

\begin{figure}
\includegraphics[width=0.5\textwidth]{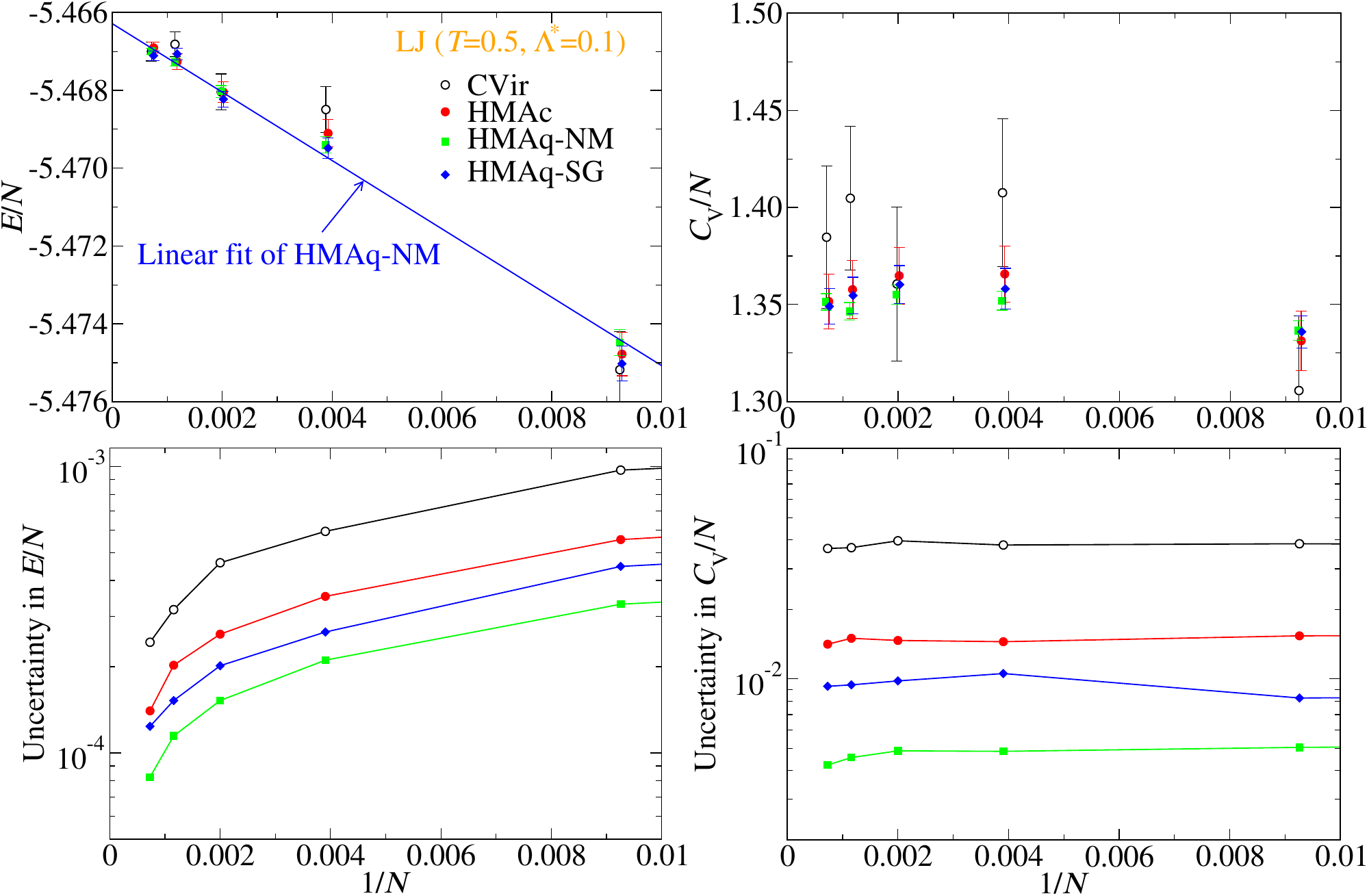}
\caption{\label{fig:FSE_N} finite size effects on the value (top) and uncertainty (bottom) of the energy (left) and heat capacity (right) of the LJ crystal at $T=0.5$ and $\Lambda^*=0.1$. The system sizes shown are (from right to left): $N=108, 256, 500, 864,$ and $1372$. The data in the top panels are shifted slightly left/right for clarity.} 
\centering
\end{figure}

\subsection{Finite Size Effects}
Path integral estimates of thermodynamic properties are subject to finite size effects, influenced by both the number of beads ($n$) and the number of atoms ($N$). However, since all estimators are based on the same primitive PI formulation, these effects should be similar across all of them. However, we focus here on the performance in terms of precision. We refer the reader to a recent work exclusively on these effects in terms of accuracy\cite{moustafa2024fse}. Figure~\ref{fig:ECv_eps2_all} shows the effect of Trotter number on the performance of the estimators in measuring the energy and heat capacity of the AO (left) and LJ (right) systems. The figure shows results for $T=0.1$, however similar behavior is observed (not shown) at other temperatures. The data are presented in terms of the parameter $\epsilon=\beta\hbar\omega/n$, which is the natural variable to describe finite size effects in quantum oscillators and crystalline systems. Note that $\omega=1.0$ for the AO system, while it is derived from the self force constant for the LJ crystal ($\omega=\sqrt{k_{\rm self}/m}$). We also present the values (top) to verify the statistical consistency among all estimators. Similar to the known behavior of the CVir estimator, the uncertainty from the HMA-based estimators does not show sensitivity to the Trotter number, especially near system sizes correspond to converged results (at about $\epsilon=0.05$).

Figure~\ref{fig:FSE_N} depicts the finite size effects in the energy (left) and heat capacity (right) of the LJ crystal in terms of the reciprocal  number of atoms. The estimators show self consistent behavior (top panels), again because they are all based on the same primitive PI formulation (eq~\ref{eq:Z_n_def}). The data then suggest that the energy has a stronger finite size effects than the heat capacity. On the other hand, the precision (lower panels) among different estimators is preserved at all system sizes and for both properties.

\section{CONCLUSIONS}
\label{sec:conclusions}
We have developed a generalized formulation for path integral estimators of total energy and heat capacity, using coordinate. The aim is to provide a framework in which precision in ensemble averages can be systematically improved, through adopting appropriate mapping fields for the given problem. The generative nature of the method allows for recovering the standard thermodynamic and CVir estimators when using the ideal gas model to define the mapping. We employ this methodology to devise novel estimators for systems with harmonic character, such as quantum oscillators and crystals. The mapping is derived from the exact solution of the HO, or the Einstein crystal for solids. In this HMA approach, two estimators emerge naturally: HMAc and HMAq, with mapping defined based on classical or quantum HO model, respectively. Although both estimators are expressed in real-space coordinates, the HMAq mapping is derived both in normal mode (HMAq-NM) and HO staging (HMAq-SG) coordinates. The computational cost for both HMAc and HMAq-SG estimators scales linearly with the Trotter number ($n$), whereas for HMAq-NM, the cost scales quadratically (or $n \log(n)$ when using discrete Fourier transform).

We assess the performance of the HMA estimators, relative to CVir, in terms of their precision (hence, CPU cost) in estimating ensemble averages of the total energy and heat capacity. The analysis is applied to a 1D (anisotropic) anharmonic oscillator and a 3D Lennard-Jones crystal, at temperature range where nuclear quantum effects are substantial. We sample coordinates using path integral molecular dynamics (PIMD) simulation, adopting the recently developed HO staging coordinates. However, the relative performance (hence conclusions) should not be sensitive to the specific sampling approach (e.g. PIMC). 

The main observation for both AO and LJ models is the systematic superior precision of the HMA estimators, over that of CVir; with the best performance obtained by HMAq-NM, followed by HMAq-SG, and then HMAc. Depending on temperature, the improvement provided by HMAq-NM is about $3-5\times$ for energy and (at least) an order-of-magnitude for the heat capacity. However, in some systems, the poor computational scaling of normal modes may offset this improvement. Although simple and low cost, the HMAc estimator only provides improvement at high temperatures and performs similarly to CVir at low temperatures. In contrast, the HMAq-SG approach delivers the closest precision to HMAq-NM. However, the performance is further improved by averaging over multiple starting beads that define staging coordinates of a given configuration. Using just a small number of starting beads ($2, 4$) significantly improves precision in both energy and heat capacity with minimal overhead cost ($<40$\%).

We investigate the effect of anisotropic anharmonicity ($k_3$) of the AO model on the precision obtained by each estimator. While the performance from CVir shows no (or negligible) dependence, the two HMAq estimators exhibit a strong dependence. As anharmonicity increases, the improvement of the HMAq estimators over CVir decreases, although they continue to demonstrate superior performance overall. By definition, the HMAq-NM estimator provides exact estimates (zero fluctuations) in the HO limit, because the mapping is already based on the HO model. The behavior of HMAc we reported above (i.e., its similarity to CVir at low-$T$ and to HMAq-NM at high-$T$) continues to hold over the entire anharmonicity range we consider. 

We also examine the effect of intrinsic quantumness ($\hbar/\sqrt{m}$) in both AO and LJ models on the improvement in precision provided by HMA over CVir estimators. For the AO model, the improvement from HMAq estimators do not show sensitivity to quantumness. On the other hand, the HMAc estimator follows the HMAq-NM behavior at high-$T$, whereas, at low-$T$, it switches performance to CVir behavior. Similar behavior is observed for the LJ case, except that the improvement from HMAq estimators show a slight decrease with increasing quantumness, especially at high temperatures.

We also examined the dependence of the precision from the HMA estimators on the Trotter number. Results from both energy and heat capacity of both models do not show noticeable dependence, a common observation with the CVir approach as well. In addition, for the LJ crystal, finite size effects in terms of number of atoms was also considered. The heat capacity (value and uncertainty) does not exhibit system-size dependence, whereas the energy show a strong dependence.  However, the improvement provided by HMA estimators over CVir shows no sensitivity over the entire range of system size.

In conclusion, due to its efficiency and simplicity, we recommend the HMAq-NM for expensive models (e.g., \textit{ab initio}) where the computational scaling with $n$ is negligible compared to the model itself. In contrast, the HMAq-SG estimator (with no or a few internal averaging over starting beads) should be the method of choice with less expensive models, such as the systems used in this work. While the HMA estimators have been described assuming forces and Hessian matrix are available, a finite difference alternative is also possible for cases where energy derivatives are not accessible. The temperature dependence of the properties being differentiated shows simple variation (nearly linear), which, in turn, yields accurate finite difference derivatives.

Since the formulation is designed for systems with harmonic character, it is (clearly) not applicable to fluids. However, the mapped averaging framework is generic, such that application to other systems is possible, given a suitable mapping reference is used. Moreover, extensions to thermoelastic properties, such as pressure and elastic constants, is straightforward. A successful example to such extension, yet for classical simulation, was recently introduced for computing elastic properties.\cite{moustafa2022cij} The efficiency of the HMA estimators to provide higher precision than the conventional centroid virial approach opens new avenue for applications to more challenging systems, such that crystals and molecular bonds treated with first principles models.

\section*{ACKNOWLEDGMENTS}
Computational resources were provided by the Center for Computational Research (CCR), University at Buffalo, NY.

\bibliography{references}

\begin{figure*}
\includegraphics[width=\textwidth]{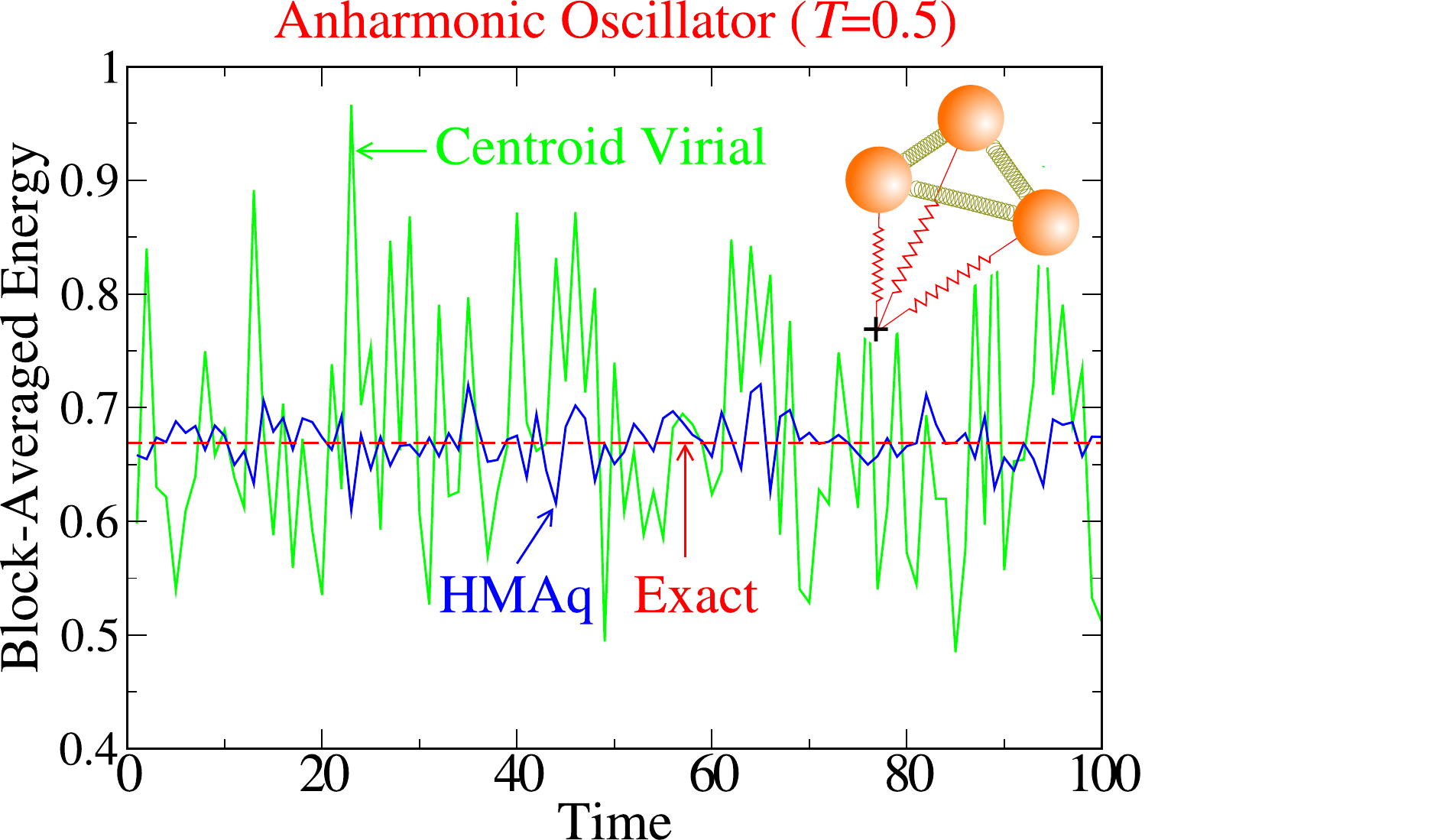}
\caption{\label{fig:toc} For table of contents graphic only}
\centering
\end{figure*}

\end{document}


\maketitle

\section{Finite difference via coordinate scaling}
\label{sec:fd_apdx}
The Eulerian version of the mapped averaging expressions (eq. 11 in the main text) relies on forces and Hessian matrix being available. However, when this is not the case, the Lagrangian alternative (eq. 7 in the main text) could be directly evaluated through finite difference schemes. For example, using two- and three-point finite difference methods for first and second derivatives, respectively, yields
\begin{subequations}

\begin{align}
\label{eq:fd_12}
\frac{D \beta V\left({\bf x}\left(\beta\right), \beta\right)}{D \beta}  &\approx  
\frac{\beta^{+} V\left({\bf x}\left(\beta^{+}\right), \beta^{+}\right) - \beta^{-} V\left({\bf x}\left(\beta^{-}\right), \beta^{-}\right)}{2\Delta \beta} \\
\frac{D^2  \beta V\left({\bf x}\left(\beta\right), \beta\right)}{D \beta^2}  &\approx  
\frac{\beta^{+} V\left({\bf x}\left(\beta^{+}\right), \beta^{+}\right) 
-2 \beta_0 V\left({\bf x}\left(\beta_0\right), \beta_0\right)
+ \beta^{-} V\left({\bf x}\left(\beta^{-}\right), \beta^{-}\right)}{\left(\Delta \beta\right)^2}
\end{align} 
\end{subequations}
where $\beta_0$ is the current value at which the derivatives are evaluated and $\beta^{\pm}\equiv \beta_0{\pm}\Delta \beta$ are the perturbed values. Although higher order schemes could be used, results show excellent agreement with the Eulerian approach. This, in turn suggests that the variation in $\beta V\left({\bf x}\left(\beta\right), \beta\right)$ with $\beta$ is nearly linear. 

For both standard and HMA estimators, we provide below prescriptions for the scaled coordinates ${\bf x}_i\equiv {\bf x}_i\left(\beta\right)$ at some $\beta$ (e.g., $\beta^{\pm}$), given an initial configuration ${\bf x}_i^0\equiv {\bf x}_i\left(\beta_0\right)$ at $\beta_0$. The derivation is based on integrating the mapping velocity $\dot{\bf x}_i$ for each case, which are given in the main text.

\subsubsection*{Thermodynamic Estimator}
Since no mapping is associated with this estimator ($\dot{\bf x}_i = 0$), the coordinates do not change with $\beta$, i.e., ${\bf x}_i={\bf x}_i^0$.

\subsubsection*{Centroid Virial Estimator}
In this case, the mapping velocity of bead and centroid coordinates are given by (see Section 2.2.2 of the main text) 
\begin{subequations}
\begin{align}
\label{eq:xdot_cvir2}
\dot{\bf x}_i &\equiv \frac{{\rm D} {\bf x}_i}{{\rm D}\beta} = \frac{1}{2\beta} \left(  {\bf x}_i - {\bf x}_{\rm c} \right) \\
\label{eq:xcdot_cvir2}
\dot{\bf x}_{\rm c} &\equiv \frac{{\rm D} {\bf x}_{\rm c}}{{\rm D}\beta} = 0
\end{align}
\end{subequations}
Since no mapping for the centroid, the ${\bf x}_{\rm c}$ coordinate does not scale with $\beta$. Hence, the solution for eq.~\ref{eq:xdot_cvir2} (main text) results in
\begin{align}
{\bf x}_i = {\bf x}_{\rm c} + \sqrt{\frac{\beta}{\beta_0}} \left(  {\bf x}_i^0 - {\bf x}_{\rm c} \right)
\end{align}
We note here that this coordinate scaling expression is the same as that derived previously by Yamamoto.\cite{yamamoto2005path} 

\subsubsection*{HMAc Estimator}
The mapping velocity for the HMAc method is given by (see Section 2.3.1 of the main text)
\begin{subequations}
\begin{align}
\label{eq:xdot_hmac2}
\dot {\bf x}_i &= \frac{1}{2\beta}\left(  {\bf x}_i - 2 {\bf x}_{\rm c} \right) \\
\label{eq:xcdot_hmac2}
\dot {\bf x}_{\rm c} &= - \frac{1}{2\beta} {\bf x}_{\rm c}
\end{align}
\end{subequations}
We start with the solution for the second equation for the centroid, which is given as ${\bf x}_{\rm c} =    \sqrt{\beta_0/\beta} \; {\bf x}_{\rm c}^0$. Then, substituting this expression into the first equation and then integrate yields
\begin{align}
{\bf x}_i = \sqrt{\frac{\beta_0}{\beta}} {\bf x}_{\rm c} + \sqrt{\frac{\beta}{\beta_0}} \left( {\bf x}_i^0 - {\bf x}_{\rm c}^0 \right)
\end{align}

\subsubsection*{HMAq Estimators}
The mapping field for HMAq estimators, in both normal mode and staging coordinates, is given by (see Section 2.3.2 of the main text)
\begin{subequations}
\begin{align}
\dot y_{k} &= g_k y_{k} \\
g_k  &= -\frac{1}{2\beta \lambda_k} \frac{\partial \left(\beta \lambda_k\right)}{\partial \beta}
\end{align}
\end{subequations}
The solution for this system is then
\begin{align}
y_k = y_k^0 \sqrt{\frac{\beta_0 \lambda^0_k}{\beta \lambda_k}}
\end{align}
For the case of HMAq-NM estimator, the $y_k$ coordinates are replaced by the normal mode coordinates $q_k$, and $\lambda_k$ parameters by $\Lambda_k$ (eq. 28 in the main text). For HMAq-NM, the $y_k$ coordinates are replaced by the staging coordinates $u_i$, and $\lambda_k$ by the staging force constants $k_i$ (see Section 2.4.2 of the main text).

Once the scaling is performed in these coordinates, transforming to Cartesian coordinates since effective potential is given in these coordinates (eq.~\ref{eq:fd_12} in the main text).

\section{HMAq-SG mapping parameters}
\label{sec:HMAq_params_apdx}
Here, we provide parameters associated with the HMAq-SG estimator (Section 2.4.2 in the main text). We start with $\gamma_i$, which can be evaluated based on eqs. 32b and 33 of the main text as
\begin{align}
\label{eq:gamma_i}
\gamma_i = \left\{
\begin{array}{ll}
\frac{1}{2\beta} -  \frac{\dot{\alpha}}{2} \left[ \coth\left(\alpha\right) + n \, \mathrm{csch}\left(n \alpha\right) \right], \; i=0 \\
\frac{1}{2\beta} -  \frac{\dot{\alpha}}{2}   \coth\left[\left(n+1-i\right)\alpha \right]  \\
+  \frac{n-i}{2} \dot{\alpha}     \sinh\left(\alpha\right) \,{\rm csch}\left[\left(n+1-i\right)\alpha\right]\,{\rm csch}\left[\left(n-i\right)\alpha\right],
 \;\; i=1,2,\dots,n-1
\end{array} 
\right.    
\end{align}
The associated derivatives are then
\begin{subequations}
\begin{align}
{\dot \gamma}_0 &= - \frac{1}{2\beta^2}
-  \frac{\ddot{\alpha}}{2} \left[ \coth\left(\alpha\right) + n \, \mathrm{csch}\left(n \alpha\right) \right]
+ \frac{\left(\dot{\alpha}\right)^2}{2} \left[{\rm csch}^2\left(\alpha\right) + n^2 \, \mathrm{csch}\left(n \alpha\right) \coth\left(n \alpha\right)\right] 
\\
{\dot \gamma}_i &= - \frac{1}{2\beta^2}   -  \frac{\ddot{\alpha}}{2}   \coth\left( \left(n+1-i\right)\alpha \right) 
+ \frac{n-i}{2} \ddot{\alpha}     \sinh\left(\alpha\right) \,{\rm csch}\left(\left(n+1-i\right)\alpha\right)\,{\rm csch}\left(\left(n-i\right)\alpha\right)
\nonumber\\
&+ \frac{\left(\dot{\alpha}\right)^2}{2} \left(n+1-i\right)  {\rm csch}^2\left( \left(n+1-i\right)\alpha \right)  + \frac{\left(\dot{\alpha}\right)^2}{2}  \left(n-i\right)  \cosh\left(\alpha\right) \,{\rm csch}\left(\left(n+1-i\right)\alpha\right)\,{\rm csch}\left(\left(n-i\right)\alpha\right) \nonumber \\
&- \frac{\left(\dot{\alpha}\right)^2}{2}  \left(n-i\right)\left(n+1-i\right)   \sinh\left(\alpha\right) \,{\rm csch}\left(\left(n+1-i\right)\alpha\right)
\coth\left(\left(n+1-i\right)\alpha\right) \,{\rm csch}\left(\left(n-i\right)\alpha\right)  \nonumber \\
&- \frac{\left(\dot{\alpha}\right)^2}{2}  \left(n-i\right)^2  \sinh\left(\alpha\right) \,{\rm csch}\left(\left(n+1-i\right)\alpha\right)\,{\rm csch}\left(\left(n-i\right)\alpha\right)\coth\left(\left(n-i\right)\alpha\right),  \;\; i=1,2,\dots,n-1
\end{align}
\end{subequations}
where $\alpha \equiv 2\sinh^{-1}\left(\frac{\epsilon}{2}\right)$, with $\epsilon \equiv \beta \hbar  \omega/n$, hence derivative are:
\begin{subequations}
\begin{align}
\dot{\alpha} &\equiv \frac{\partial \alpha}{\partial \beta} 
= \frac{\hbar \omega}{n} \left( 1+ \frac{\epsilon^2}{4} \right)^{-1/2} \\
\ddot{\alpha} &\equiv \frac{\partial^2 \alpha}{\partial \beta^2} 
= - \frac{\epsilon}{4}\left(\frac{\hbar \omega}{n}\right)^2 \left( 1+ \frac{\epsilon^2}{4} \right)^{-3/2} 
\end{align}
\end{subequations}

The $A_i$ and $B_i$, and associated derivatives, are given by:
\begin{subequations}
\begin{align}
A_i &\equiv \sinh\left(\alpha\right) \,{\rm csch}\left(\left(n+1-i\right) \alpha\right) \\
\dot{A}_i &= \dot{\alpha} {\rm csch}
\left( \left(n+1-i\right) \alpha\right)  \left[\cosh\left(\alpha\right)  - \left(n+1-i\right)  \sinh\left(\alpha\right) \coth\left(\left(n+1-i\right) \alpha\right) 
\right] \\
\ddot{A}_i &= \frac{\ddot{\alpha}}{\dot{\alpha}} \dot{A}_i + \dot{\alpha}^2 {\rm csch}\left(\left(n+1-i\right)\alpha\right)  \nonumber \\
&\times \left[ -2\left(n+1-i\right) \cosh\left(\alpha\right) \coth\left(\left(n+1-i\right)\alpha\right) +   \sinh\left(\alpha\right) 
+  \frac{1}{2}\left(n+1-i\right)^2 \sinh\left(\alpha\right) \, {\rm csch}^2\left(\left(n+1-i\right)\alpha\right)
\left(3 + \cosh\left(2\left(n+1-i\right)\alpha\right)\right)
\right]
\end{align}
\end{subequations}
and
\begin{subequations}
\begin{align}
 B_i &\equiv \sinh\left(\left(n-i\right)\alpha\right) \,{\rm csch}\left(\left(n+1-i\right) \alpha\right) \\
\dot{B}_i &= \dot{\alpha}
{\rm csch}\left(\left(n+1-i\right) \alpha\right) 
\left[
\left(n-i\right) \cosh\left(\left(n-i\right)\alpha\right)
- \left(n+1-i\right)  \sinh\left(\left(n-i\right)\alpha\right) \coth\left(\left(n+1-i\right) \alpha\right)
\right] \\
\ddot{B}_i &= \frac{\ddot{\alpha}}{\dot{\alpha}} \dot{B}_i + \dot{\alpha}^2  {\rm csch}\left(\left(n+1-i\right) \alpha\right) \nonumber \\
&\times \Big[
 -2\left(n+1-i\right)\left(n-i\right) \cosh\left(\left(n-i\right)\alpha\right) \coth\left(\left(n+1-i\right)\alpha\right) 
 + \left(n-i\right)^2 \sinh\left(\left(n-i\right)\alpha\right)
  \nonumber \\
&+ 
\frac{1}{2}\left(n+1-i\right)^2 \sinh\left(\left(n-i\right)\alpha\right)  \, {\rm csch}^2\left(\left(n+1-i\right)\alpha\right)
\left( 3 + \cosh\left(2\left(n+1-i\right)\alpha\right)  \right)
\Big]   
\end{align}
\end{subequations}

\section{Center of mass correction for HMA estimators}
\label{sec:com}
Coordinate mapping in HMA estimators depends on the HO model (or Einstein crystal for solids), which is a bound potential that depend on absolute coordinates. However, for translationally invariant (unbound) models, such as Lennard-Jones, the potential energy depends only on relative coordinates, such as those measured with respect to the center of mass (COM). Therefore, for the HO-based mapping to describe unbound models, we need to remove the COM contribution of the mapping velocity.  Since standard potentials depend already on unbound models (ideal gas), such step is not needed. We emphasize that there is no COM contribution associated with bound models such as quantum oscillators and bonds. Therefore, for our case, we only considered this contribution to the LJ crystal, and not to the AO system.

The COM position of a system of $N$ atoms is defined as
\begin{align}
\label{eq:com_vel}
{\bf r}_{\rm com}    = \frac{1}{Nn} \sum_{j=1}^{N} \sum_{i=0}^{n-1} {\bf r}_{i}^{j}= \frac{1}{N} \sum_{j=1}^{N}  {\bf r}_{\rm c}^j \end{align}
where ${\bf r}_i^j$ is the position of bead $i$, inside a ring-polymer $j$, and ${\bf r}_{\rm c}^j$ is the associated centroid coordinate. Then, this contribution needs to be subtracted from the divergence term of the mapped averaging formulas, 
\begin{align}
\sum_{j=1}^{N} \sum_{i=0}^{n-1} \nabla_{ij}  \cdot  \dot {\bf r}_i^j  
\to
\sum_{j=1}^{N} \sum_{i=0}^{n-1} \nabla_{ij}  \cdot \left( \dot {\bf r}_i^j - \dot {\bf r}_{\rm com} \right)
\end{align}
This is a compact form, in which $\dot {\bf x}$ represents a mapping velocity vector of all $dNn$ degrees of freedom and $\dot {\bf x}_{\rm com}$ is a $dNn$-length vector made from $Nn$ blocks of \textit{identical} COM vectors of length $d$, with each component defined by eq~\ref{eq:com_vel}. 

Clearly, the mapping velocity enters into terms other than the divergence (eq. 11). However, due to the translational invariance nature of the unbound models, the sum over all forces and Hessian is zero. Since the COM mapping velocity is a constant subtracted from from the mapping velocity of each bead, its contribution will get cancelled.    

Below, we evaluate the COM contribution for both the HMAc and HMAq estimators. In all cases, we find that the COM contribution is merely a finite-size effect, vanishing as $1/N$ in the thermodynamic limit. Therefore, the same thermodynamic limit of properties can be obtained without including these contributions. However, they are useful when comparing HMA estimators against standard estimators, which is the case in this study with LJ crystal.

\subsection*{HMAc case}
For HMAc estimator (Section 2.3.1 of the main text), the centroid mapping velocity is given by
\begin{align}
\label{eq:xcdot_hmac_si}
\dot {\bf r}_{\rm c}^j = -\frac{1}{2\beta} {\bf r}_{\rm c}^j   = -\frac{1}{2\beta n} \sum_{i=0}^{n-1} {\bf r}_i^j
\end{align}
Hence, according to eq~\ref{eq:com_vel}, the COM mapping velocity is
\begin{align}
\dot {\bf r}_{{\rm com}}  &=  -\frac{1}{2\beta N n} \sum_{i=0}^{n-1} \sum_{j=1}^N {\bf r}_i^j
\end{align}
The associated divergence and its derivative are
\begin{align}
\sum_{j=1}^{N} \sum_{i=0}^{n-1} \nabla_{ij}  \cdot  \dot {\bf r}_{\rm com}   &= -\frac{d}{2\beta}
\end{align}
The reference term in the HMAc expressions of energy and heat capacity are then
\begin{subequations}
\begin{align}
E^{\rm HOc} &\to \frac{dN}{\beta}    - \frac{d}{2\beta} \\
\frac{C^{\rm HOc}_{\rm V}}{k_{\rm B}\beta^2} &\to \frac{dN}{\beta^2}    - \frac{d}{2\beta^2}
\end{align}
\end{subequations}

\subsection*{HMAq-NM case}
We will consider here the case of HMAq-NM (same analysis can be applied to HMAq-SG). In the NM coordinates, the $x$ component of the centroid position associated with atom $j$ is given by $x^j_{\rm c} = q_0^j$ (see Section 2.4.1). Hence, according to eqs. 25a and 31, the centroid mapping velocity is
\begin{align}
\dot x_{\rm c}^j  = \dot q_{0,j} =  -\frac{1}{2\beta} q_{0,j}  =   -\frac{1}{2\beta} x_{\rm c}^j
\end{align}
This is identical to the HMAc case (eq~\ref{eq:xcdot_hmac_si}), hence same COM terms need to be subtracted from HMAq-NM expressions (eq. 26),
\begin{subequations}
\begin{align}
E^{\rm HMAq} &\to E^{\rm HMAq}     - \frac{d}{2\beta} \\
\frac{C^{\rm HMAq}_{\rm V}}{k_{\rm B}\beta^2} &\to \frac{C^{\rm HMAq}_{\rm V}}{k_{\rm B}\beta^2}    - \frac{d}{2\beta^2}
\end{align}
\end{subequations}

\subsection*{HMAq-SG case}
Unlike the HMAq-NM case, the centroid mode does not correspond to any particular staging coordinates; rather, all coordinates contribute to it. In this case, it is more convenient to define the $x$ component of the COM as
\begin{eqnarray}
\label{eq:r_com_sg}
x_{\rm com} = \frac{1}{Nn} \sum_{k=1}^{N}  J {\bf x}^{k}
\end{eqnarray}
where $J$ is $1\times dn$ vector of ones and ${\bf x}^k$ is a $n\times 1$ vector of the $x$ component of all beads coordinates associated with the atom/ring-polymer $k$. The staging coordinates are given as ${\bf u}^k = L {\bf x}^k$, where $L$ is the transformation matrix, with components given by eq. 23c. The associated mapping velocity is $\dot{\bf u}^{j} = \Gamma {\bf u}^{j}$, where $\Gamma$ is a diagonal matrix with components given by $\gamma_i$ (eq~\ref{eq:gamma_i}). The associated mapping velocity in Cartesian coordinates is then $\dot{\bf x}^k = M {\bf x}^k
$, where $M \equiv L^{-1}\Gamma L- L^{-1}\dot L$ is a $n\times n$. The mapping velocity of COM (eq~\ref{eq:r_com_sg}) is then
\begin{eqnarray}
\label{eq:rdot_com_sg}
\dot x_{\rm com} = \frac{1}{Nn} \sum_{k=1}^{N}  J M {\bf x}^{k}
\end{eqnarray}
Since $x_i^k$ represents the $k$ component of ${\bf x}^k$, the $x$ contribution of the divergence is
\begin{align}
\sum_{j=1}^{N} \sum_{i=0}^{n-1} \frac{\partial  \dot x_{\rm com}}{\partial x_i^j}   
= \frac{1}{Nn}\sum_{j=1}^{N}\sum_{i=0}^{n-1} JM   \frac{\partial  {\bf x}^{j}}{\partial  x_i^j} = \frac{1}{Nn}\sum_{j=1}^{N}\sum_{i=0}^{n-1}  \sum_{i'=0}^{n-1} M_{i',i} = \frac{1}{n} \sum_{i,i'} M_{i',i}
\end{align}
where we used the fact that $\left(\frac{\partial  {\bf x}^{j}}{\partial x_i^j} \right)_k=\delta_{k,i}$ and $M$ matrix is the same for all ring polymers (i.e., independent of $j$). The full divergence term in 3D is then $\frac{d}{n} \sum_{ij} M_{ij}$. This COM contribution needs to be subtracted from the HMAq-SG estimator expressions (eq. 26),
\begin{subequations}
\begin{align}
E^{\rm HMAq} &\to E^{\rm HMAq}     - \frac{d}{n} \sum_{ij} M_{ij}  \\
\frac{C^{\rm HMAq}_{\rm V}}{k_{\rm B}\beta^2} &\to \frac{C^{\rm HMAq}_{\rm V}}{k_{\rm B}\beta^2}    -  \frac{d}{n} \sum_{ij} \dot M_{ij}
\end{align}
\end{subequations}
where $\dot M$ is the $\beta$-derivative of $M$.

Unlike previous cases, the COM expression of HMAq-SG is more complicated to evaluate analytically in a closed form. However, numerical evaluations indicate that the values are not far from those of HMAc and HMAq-NM; hence, we use them for the HMAq-SG case as well. Although approximate, but as mentioned earlier the COM is merely a finite-size effect contribution, which does not affect extrapolated thermodynamic limit estimates.

\section{PIMD step size effects}
\label{sec:dt_si}
\begin{figure*}
\centering
\subfloat[\centering AO ($\Lambda^*=1.0$)]{{\includegraphics[width=0.48\textwidth]{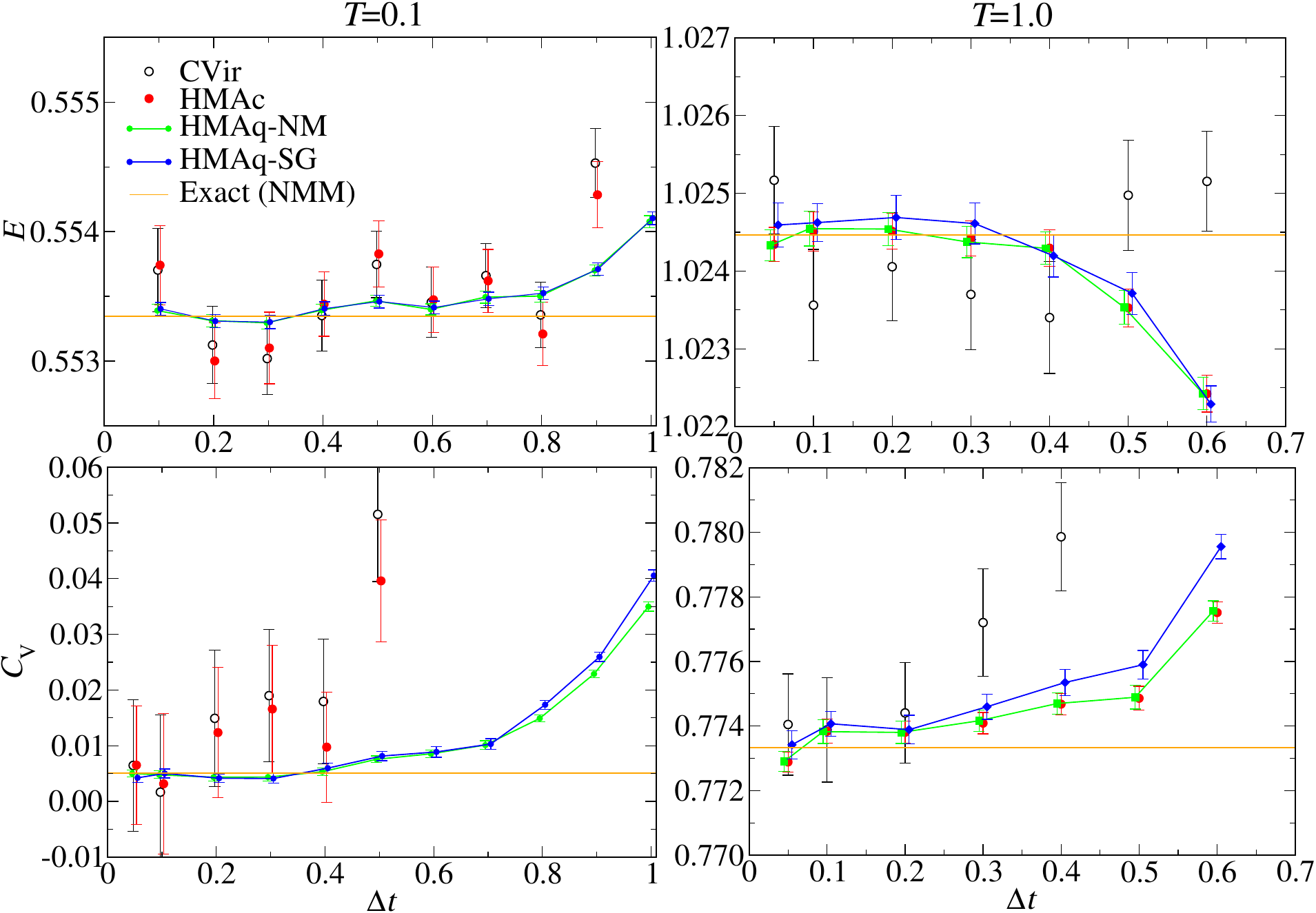} }}%
\quad
\subfloat[\centering LJ ($\Lambda^*=0.1$)]{{\includegraphics[width=0.48\textwidth]{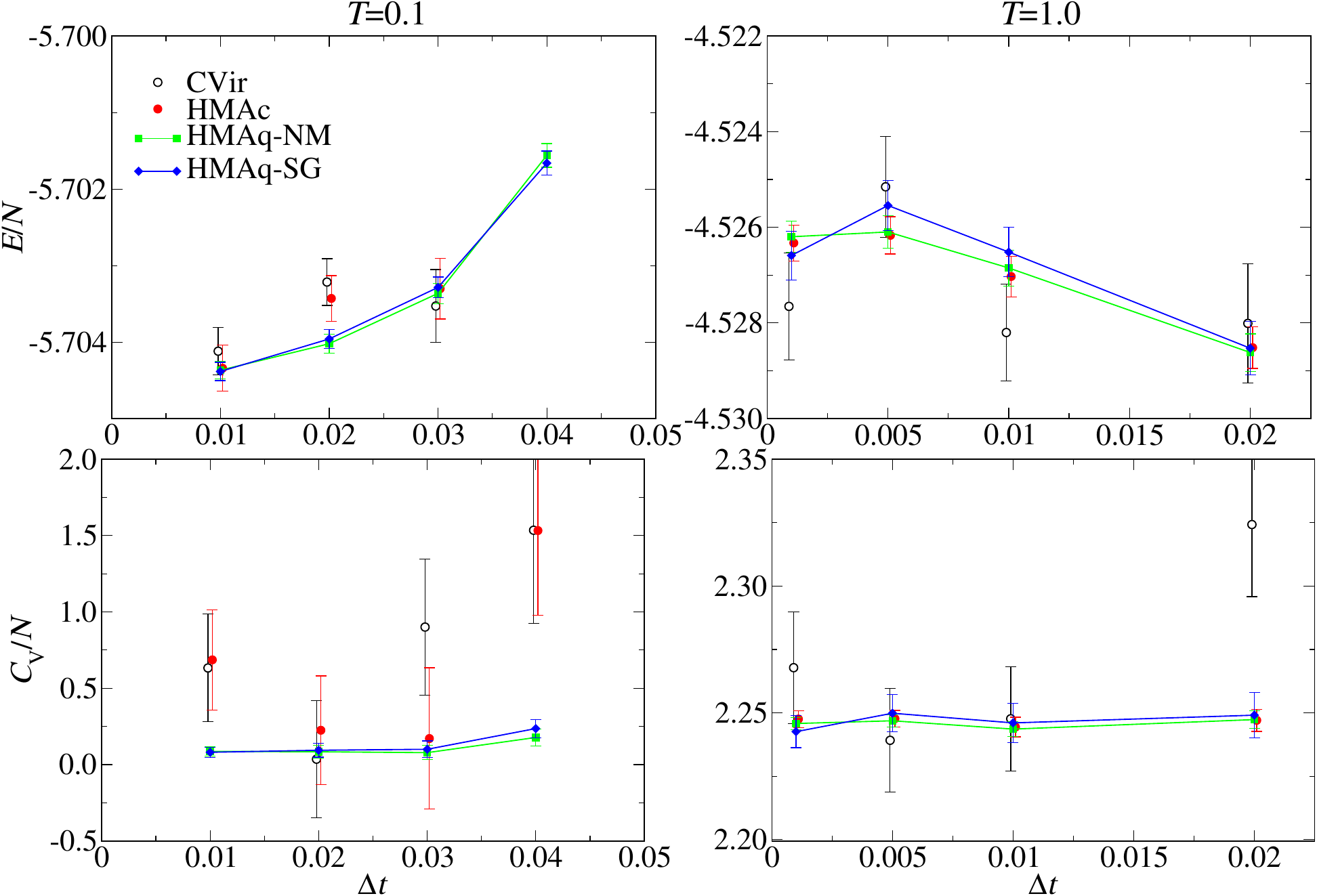} }}%
\caption{\label{fig:ECv_dt_all} Dependence of the assemble averages of the energy (top) and heat capacity (bottom) on the PIMD time step size ($\Delta t$) for the AO (left) and LJ (right) models, at low ($T=0.1$) and high ($T1.0$) temperatures. The number of steps $N_{\rm steps}$ is chosen such that the simulation time ($t_{\rm sim} = N_{\rm steps} \Delta t$) is kept fixed for all points, with $t_{\rm sim}=10^6$ (AO) and $10^3$ (LJ). For AO, the orange horizontal line represent the exact value as computed using the NMM method. Presented results correspond to stable integrator (simulations failed for larger $\Delta t$ values). The data are shifted slightly left/right for clarity.}
\end{figure*}

We investigate the accuracy of estimators in terms of convergence of properties with respect to the PIMD time step size ($\Delta t$). Figure~\ref{fig:ECv_dt_all} presents the convergence of total energy (top) and heat capacity (bottom) of the AO (left) and LJ (right) models, both at low ($T=0.1$) and high ($T=1.0$) temperatures. To reduce variations in the statistical uncertainty (error bars) across different time steps, we choose the number of PIMD steps such that the simulation time remains constant ($t_{\rm sim}=N_{\rm steps}\Delta t$). For the AO model, we include the exact values using the numerical matrix multiplication (NMM) method.\cite{NMM1983,NMM2001} Statistically, the convergence rate of the total energy of both models appears to be similar, regardless of the estimator type. For example, values of $\Delta t=0.5$ and $0.3$ seem to provide statistically converged results for the AO model at the low and high temperatures, respectively, while the LJ energy appears to converge at a common value of $\Delta t=0.01$.  In contrast, for the heat capacity, the HMAq estimators show faster convergence rate than the CVir estimator. This is especially the case with the LJ model, where results appear to be nearly statistically flat over the entire domain of integrator stability.

\bibliography{references}